\documentclass[]{spie}  

\usepackage{amsmath,amsfonts,amssymb}
\usepackage{graphicx}
\usepackage[colorlinks=true, allcolors=blue]{hyperref}
\usepackage{float}

\title{Use of 3D printing in astronomical mirror fabrication}

\author[a]{M\'elanie Roulet}
\author[b]{Carolyn Atkins}
\author[a]{Emmanuel Hugot}
\author[c]{Robert Snell}
\author[d]{Bart van de Vorst}
\author[b]{Katherine Morris}
\author[a]{Michel Marcos}
\author[c]{Iain Todd}
\author[b]{Christopher Miller}
\author[d]{Joris Dufils}
\author[e]{Szigfrid Farkas}
\author[e]{Gyorgy Mezo}
\author[f]{Fabio Tenegi}
\author[f]{Afrodisio Vega-Moreno}
\author[b]{Hermine Schnetler}
\affil[a]{Aix Marseille Univ, CNRS, CNES, LAM, Marseille, France}
\affil[b]{UK Astronomy Technology Centre, Royal Observatory, Blackford Hill, Edinburgh, EH9 3HJ, UK}
\affil[c]{Department of Materials Science and Engineering, University of Sheffield, Mappin Street, Sheffield, S1 3JD, UK}
\affil[d]{L.T.G. van de Vorst TNO, 5656 AE Eindhoven, the Netherlands}
\affil[e]{Konkoly Observatory, Research Centre for Astronomy and Earth Sciences, MTA Centre for Excellence, 1121 Konkoly-Thege Miklós út 15-17, Budapest, Hungary}
\affil[f]{Instituto de Astrofísica de Canarias (IAC), 38200, La Laguna, (S/C Tenerife), Spain}

\authorinfo{melanie.roulet@lam.fr}

\begin{document} 
\maketitle

\begin{abstract}

In this paper we are exploring the possibilities of 3D printing in the fabrication of mirrors for astronomy. Taking the advantages of 3D printing to solve the existing problems caused by traditional manufacturing, two proof-of-concept mirror fabrication strategies are investigated in this paper. The first concept is a deformable mirror with embedded actuator supports system to minimise errors caused by the bonding interfaces during mirror assembly. The second concept is the adaption of the Stress Mirror Polishing (SMP) technique to a variety of mirror shapes by implemented a printed thickness distribution on the back side of the mirror. Design investigations and prototypes plans are presented for both studies. 
\end{abstract}

\keywords{3D printing, mirror fabrication, polishing techniques, finite element analysis, additive manufacturing, deformable mirror, Off axis parabolas, stress polishing}

\section{INTRODUCTION}
First created for commercial industry, 3D printing is increasingly used in leading-edge fields such as space instrumentation. Optical components and telescope structures are now under investigation to be printed. 3D printing provides a new vision for optics fabrication due to the large scale of techniques and materials available. In the field of space instrumentation, the challenge is to improve crucial parameters such as bulkiness and weight whilst maintaining the optical performance as surface quality. The use of 3D printing to create novel lightweight structures is a new and developing field. Combined with the work of topology optimisations and lattices several prototypes have been proposed by J. Mici et al. (2015) \cite{10.1117/12.2188533}, C.Atkins et al. (2019) \cite{10.1117/12.2528119} and E.Hilpert et al. (2019) \cite{10.1117/1.OE.58.9.092613}. All the prototypes have been printed in metals and shows very good surface quality after polishing or diamond turning. 

3D printing can also print ceramic material using stereolithography techniques. In traditional manufacture ceramics are often used in mirror fabrication however, there are few investigations in the use of 3D printed ceramics. Stereotlithography uses photopolymerisation reaction on a liquid resin to build the object layer by layer. 
In contrast, 3D printed metals utilise a thermal reaction between a laser (or electron beam) and metallic powder. In this paper we are interested in the printing of ceramics compared to metal printing and traditional manufacture, by highlighting its behaviour in terms of polishing and surface quality after printing. 

In the exploration of 3D printing benefits we focus on two parameters: the ability to remove interfaces and simplify assembly of complex structures in manufacture; and the removal of design constraints linked to traditional manufacture, which opens a new spectrum of creativity for mirror design. In this paper two experiments on the application of 3D printing for astronomical mirrors are explored. 

The first parameter is explored in the context of a deformable mirror, where the actuator-to-mirror interfaces increase the manufacture risk and, in addition, the error budget. 3D printing provides technical solutions to delete interfaces, such as bonding, by printing the mirror and the actuators supports in one step. Therefore, we present mirrors that have been printed with ribs structure as actuator interfaces to investigate the impact of the polishing on the surface quality. Circular and honeycomb support structures have been added to reduce the print through of the rib structure during polishing. In addition, 3D printing provides new ceramics and metals with new properties whose behaviour haven’t been fully characterised under polishing. In this study, both alumina and titanium have been printed and investigated, surface quality results are compared.

For the second parameter, we investigated an innovative manufacturing process that combines 3D printing and stress polishing, to create a thickness distribution capable of producing off-axis parabola proﬁle. Three prototypes have been produced: one fully Zerodur done by traditional manufacturing; one fully printed in ceramic; and one assembly of a Zerodur blank glued to printed ceramic thickness distribution. The designs have been simulated with Finite Element Analysis (FEA) and optimised. Then physical deformations have been applied front of an interferometer. We present the comparison between the measured phase maps and the ones expected from simulation.

\section{Deformable Mirror}

\subsection{Objectives}
This study forms part of the OPTICON Additive Astronomy Integrated component Manufacturing $A^2IM$ project. The work-package 5.1.2 is dedicated to the development of deformable mirror manufactured with 3D printing. The objective is to create a very thin deformable mirror with embedded actuator supports. In the continuity of Freeform Active Mirror Experiment (FAME) \cite{10.1117/1.OE.53.3.031311}, the idea is to build a very thin deformable mirror with integrated actuators supports to remove of the assembly steps, as bonding the actuator supports after polishing degraded the surface quality. To expand the study, we are interested in other actuator patterns with thin optical surface that could be good candidate to be printed. We were inspired by the rib structure design of J. Steeves et al. (2018) presented at SPIE Astronomical Telescopes + Instrumentation conference in 2018 \cite{10.1117/12.2312645} and with reference to the work by W. Xin Lu et al (2014) \cite{10.1117/12.2069560}. The actuation design is based upon triangular rib structures and it is the imprint of these structures during manufacture that we wish to quantify. In this section is presented the design study to reduce the imprint of the structure by adding support patterns and the prototyping plan.

\subsection{Design and Finite Element Analysis}
\subsubsection{Initial design}
\begin{figure}[htbp]
    \centering
    \includegraphics[width=.45\textwidth]{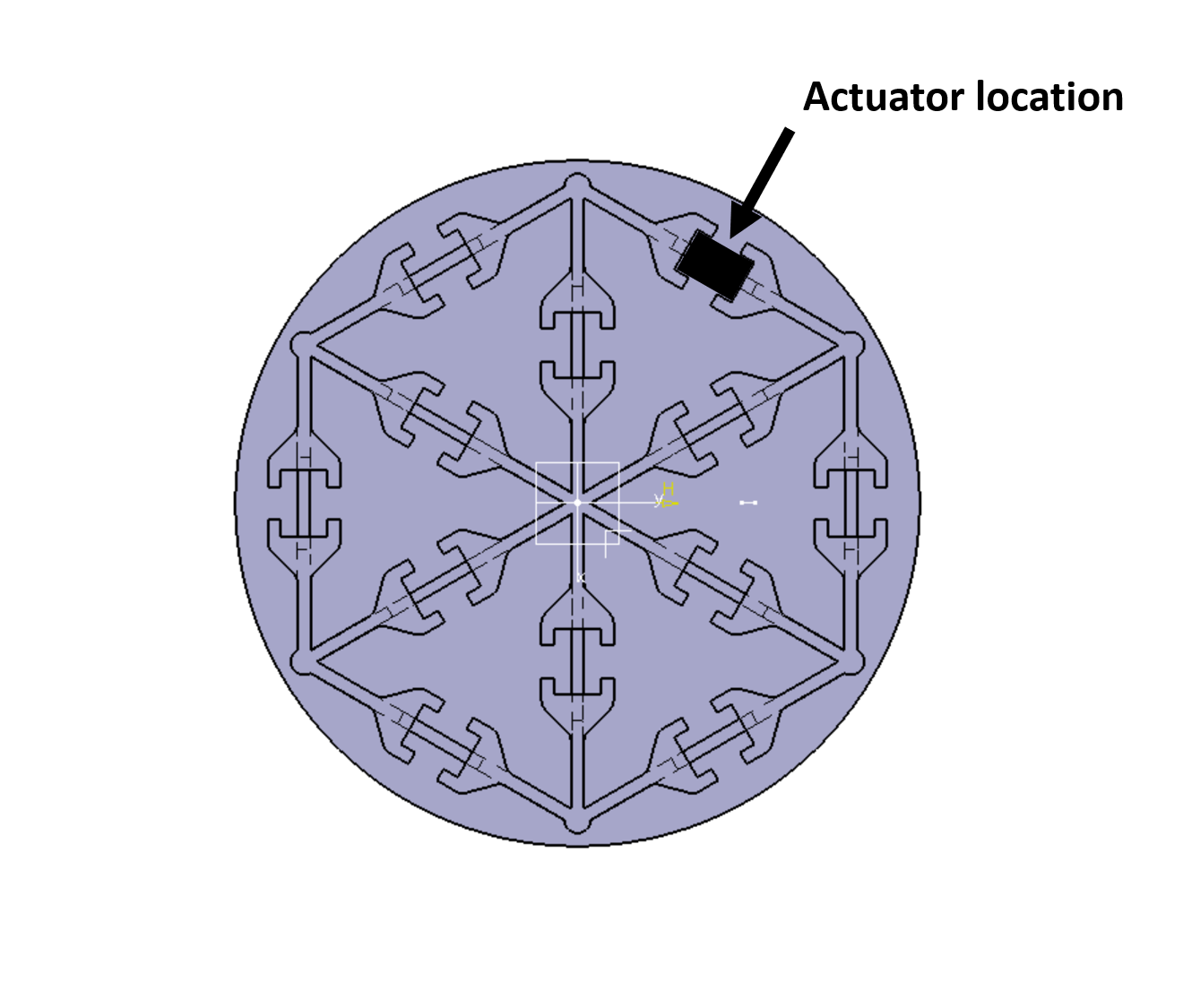}
    \includegraphics[width=.45\textwidth]{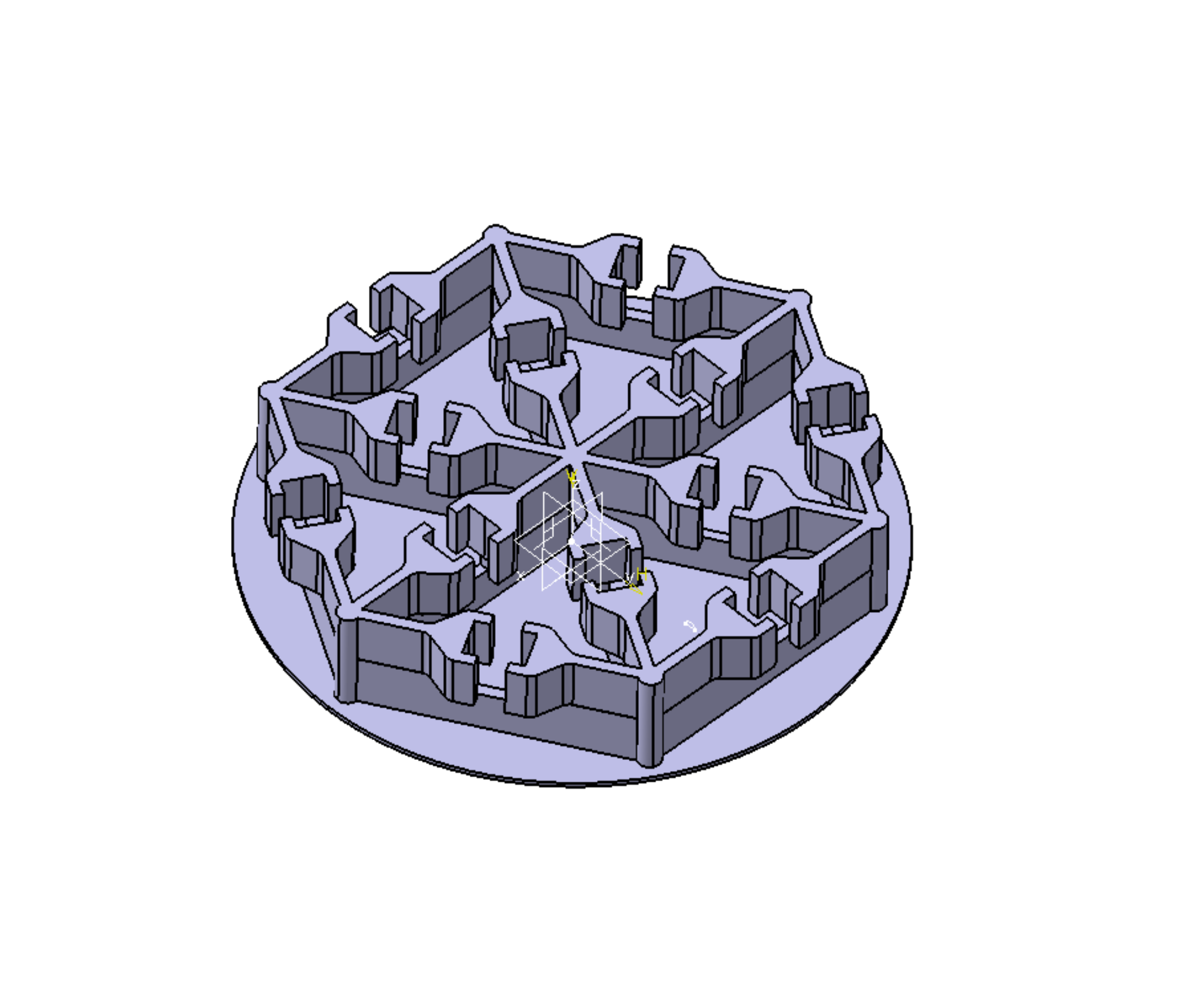}
    \caption{CAD view of the initial design with the embedded actuation structure.}
    \label{fig:InitialDesign}
\end{figure}

Figure \ref{fig:InitialDesign} presents the initial design with the back side view of the actuation structure. The model is 51 mm in diameter and the thickness of the optical surface is 0.4 mm. The total height of the mirror is 9.5 mm.

\begin{figure}[htbp]
    \centering
    \includegraphics[width=.5\textwidth]{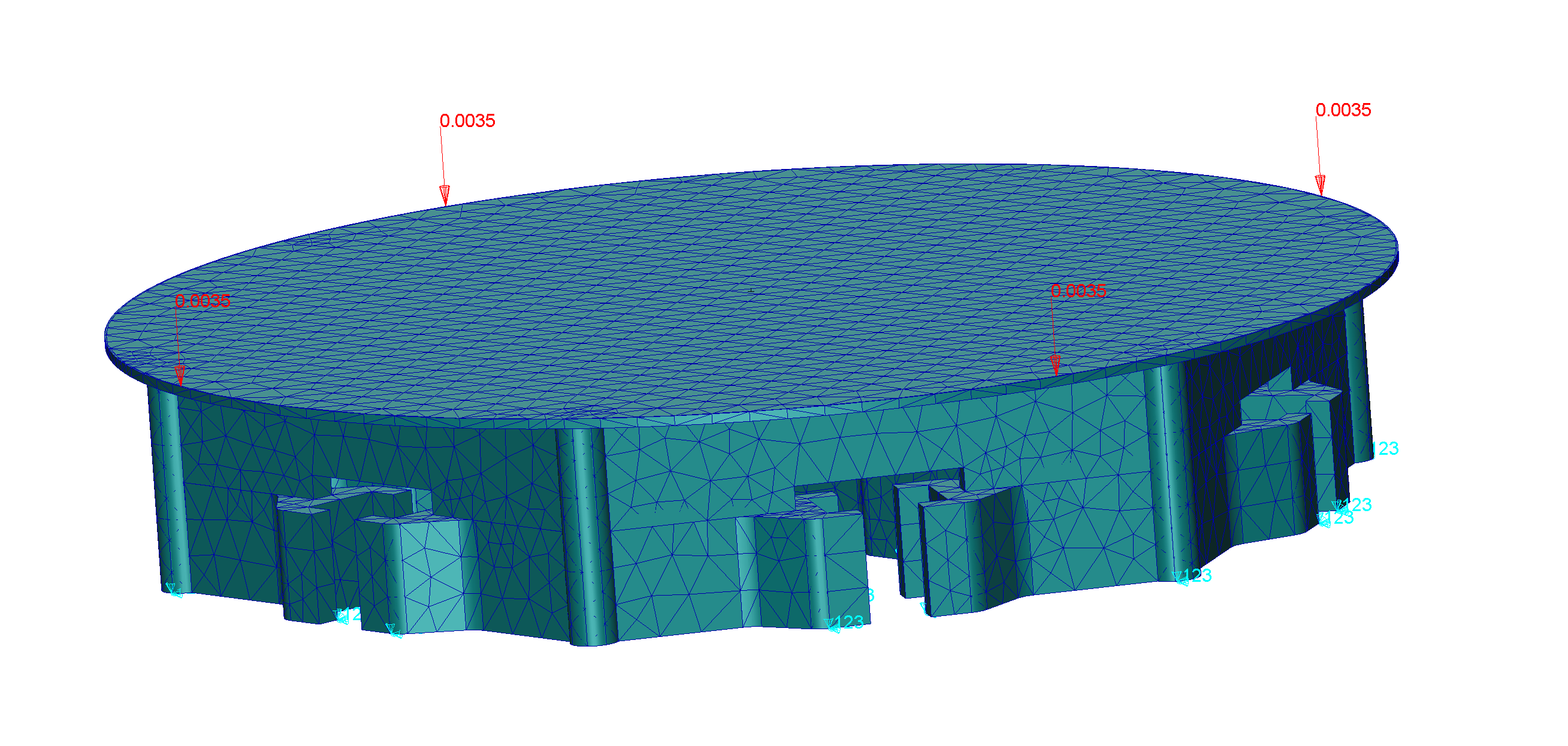}
    \caption{FEA model 3D view with the uniform pressure apply on the optical surface in red and the fixed base in blue}
    \label{fig:FEA}
\end{figure}

The polishing pressure normal to the mirror surface is simulated with Finite Element Analysis (FEA) by applying a uniform pressure of 3.5 MPa on the optical surface fixing the base, as shown in Figure \ref{fig:FEA}. The model is meshed with tetrahedral elements, aluminium material is used in the simulation. 

\begin{figure}[htbp]
    \centering
    \includegraphics[width=.4\textwidth]{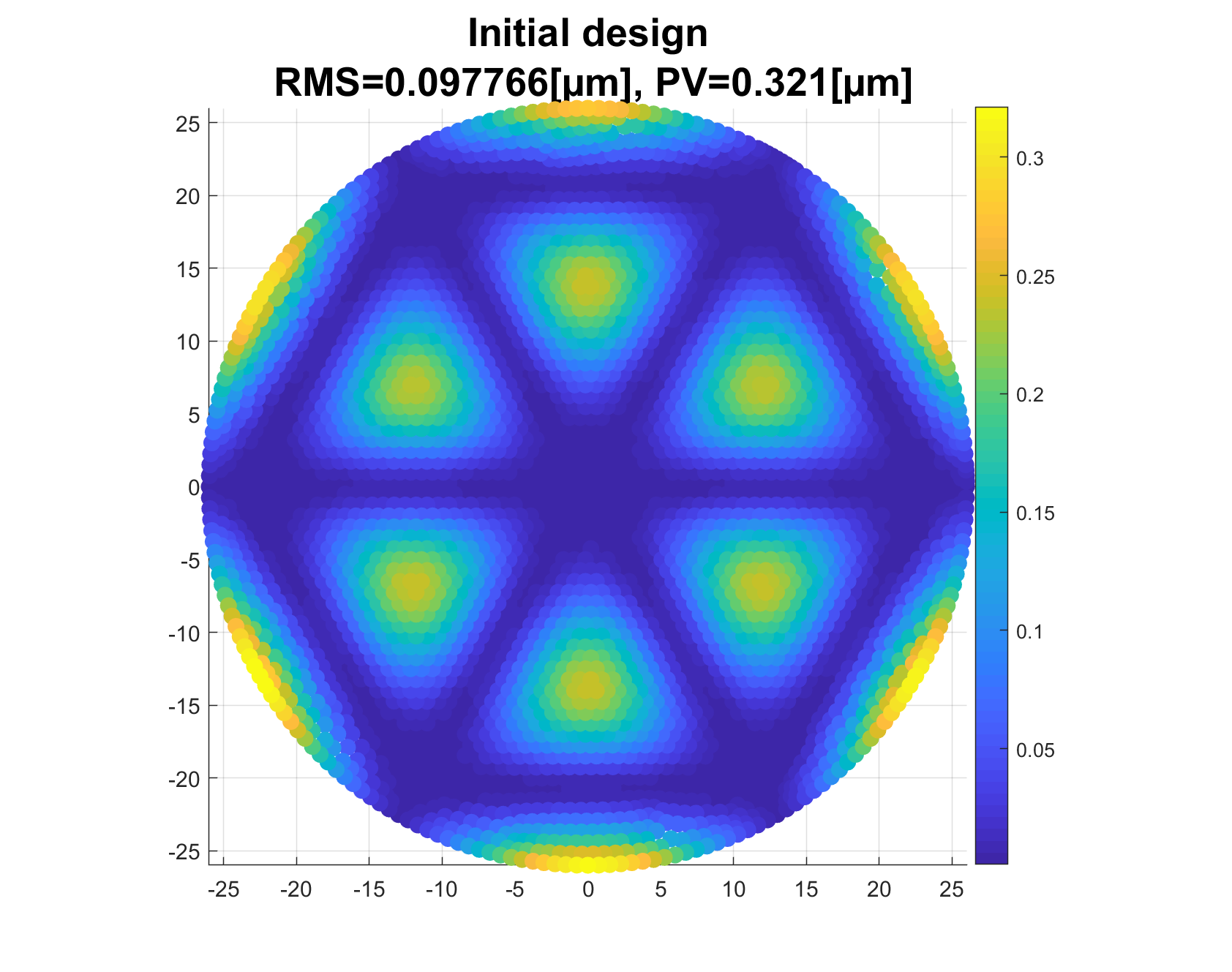}
    \caption{FEA displacements on the optical surface obtained on the initial mirror design with 0.4 mm surface thickness subject to polishing pressure of 3.5 MPa}
    \label{fig:FEAIni}
\end{figure}

Simulated displacements on the optical surface of the initial design are shown in Figure \ref{fig:FEAIni},  we can clearly see the imprint of the rib structure in blue, where the surface doesn't move. The large bending on the edges is caused by the external hexagonal structure of the actuation supports while the optical surface is circular. There are also valleys inside the triangle structure resulting of the lack of supports.

\subsubsection{New designs}
To avoid the imprint of the rib structure and the edge bending we decided to implement additional patterns in order to support the optical surface during the polishing. To compare the results on the optical surface displacements, the patterns should have a similar volume.

\begin{figure}[htbp]
    \centering
    \includegraphics[width=.45\textwidth]{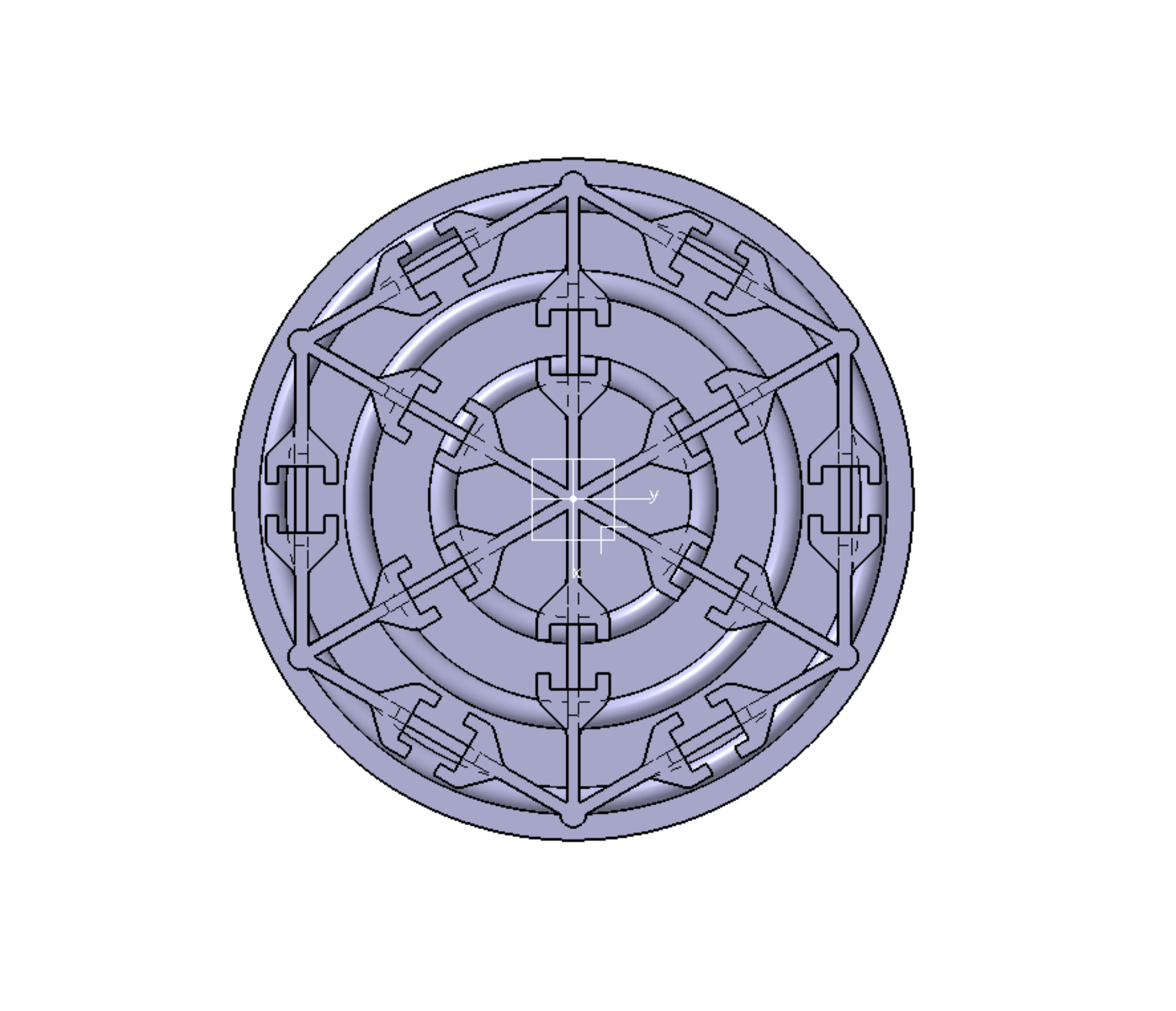}
    \includegraphics[width=.45\textwidth]{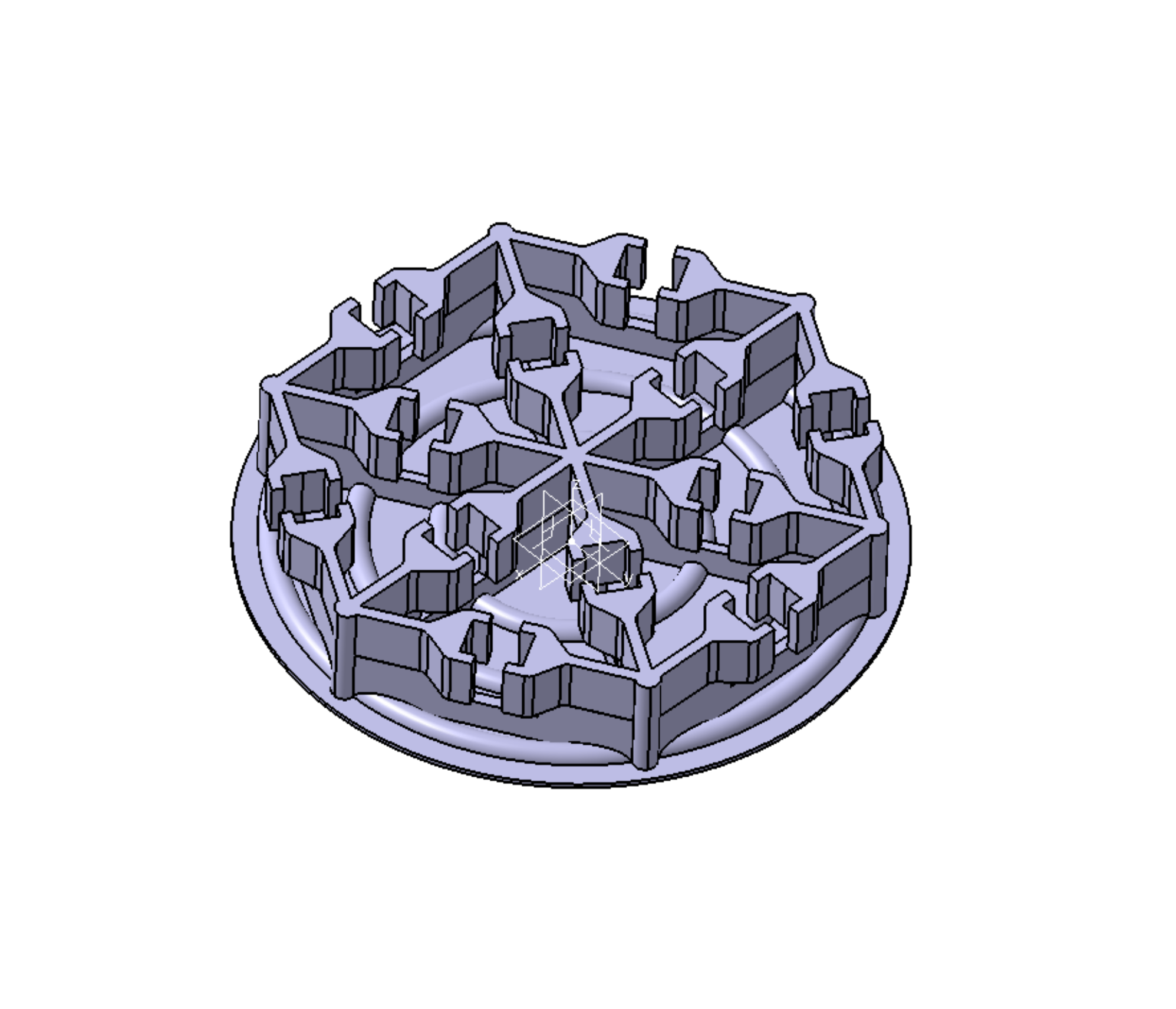}
    \caption{CAD view of the circle design }
    \label{fig:CircleDesign}
\end{figure}

We first implemented a circular pattern composed by three concentric circles in Figure \ref{fig:CircleDesign}. This structure is inspired from topology optimisation results done by C. Atkins et al (2017) \cite{10.1117/12.2274011}. The optimisation performed aimed to lightweight a small circular mirror under a uniform polishing pressure. The result is that the internal lightweighting consists of three concentric rings. To incorporate this design optimisation, we choose three concentric circles with 2 mm diameter and which are located at 10 mm, 16.5 mm, and 23 mm from the centre. This represent an additional volume of 478.2 $mm^3$.

\begin{figure}[htbp]
    \centering
    \includegraphics[width=.45\textwidth]{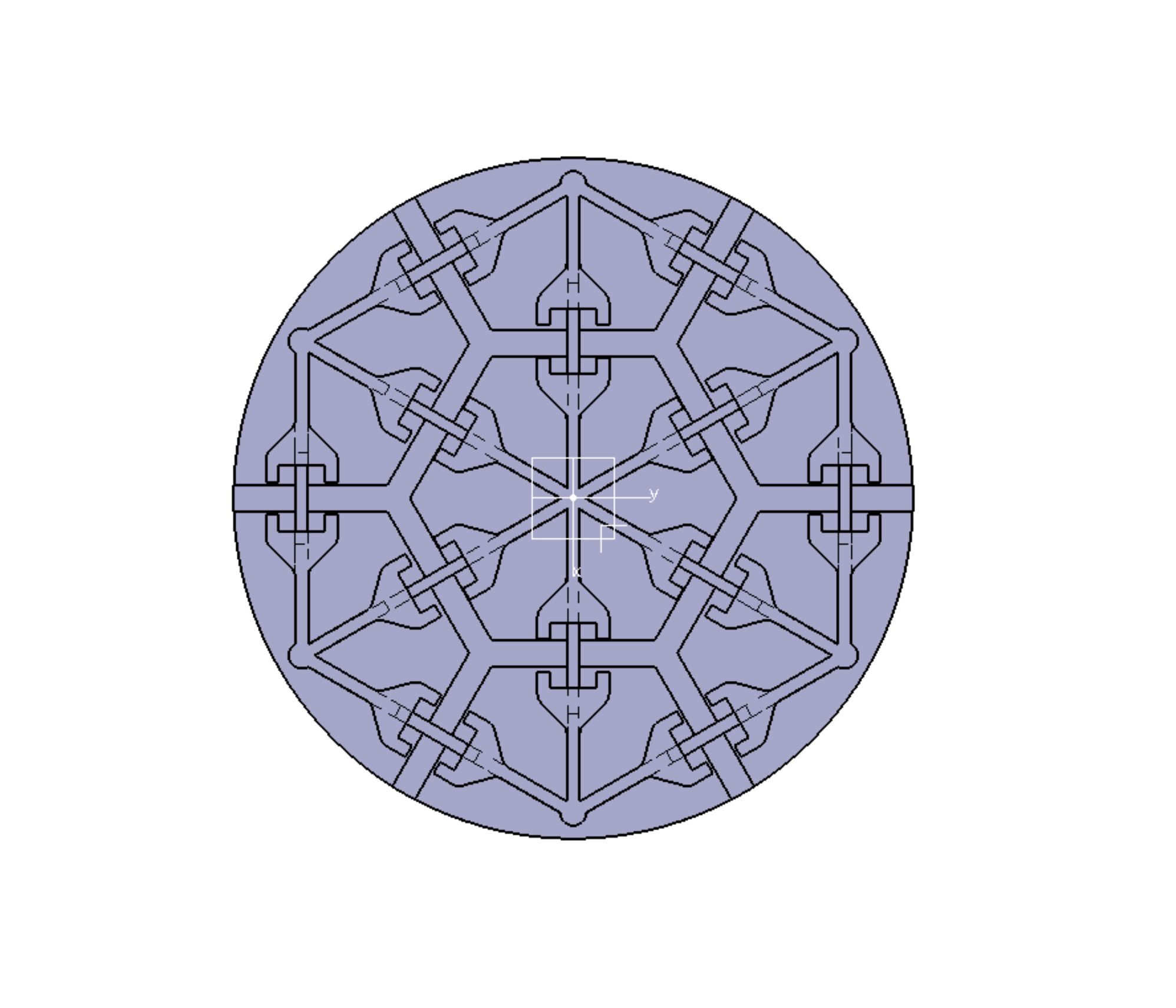}
    \includegraphics[width=.45\textwidth]{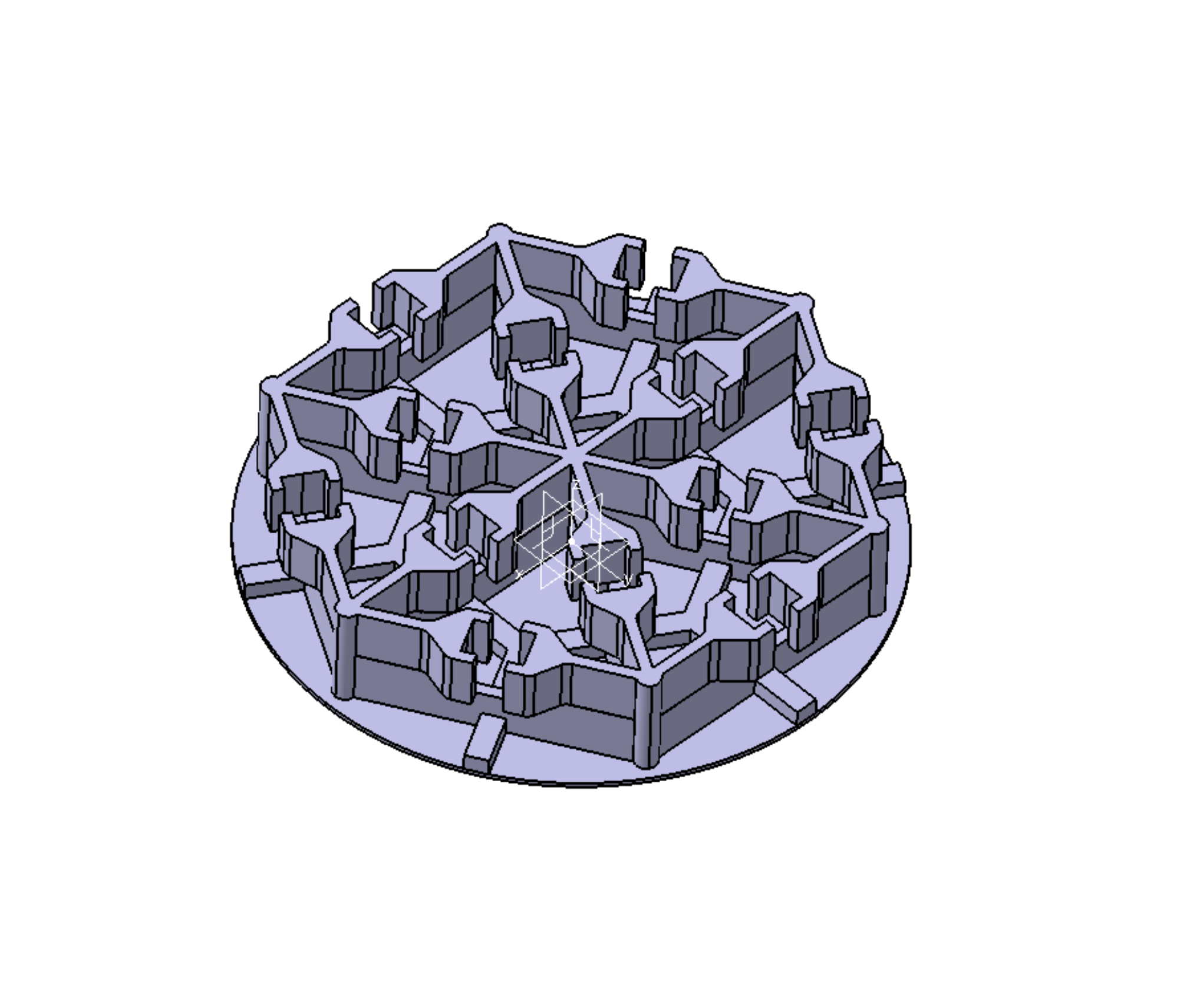}
    \caption{CAD view of the honeycomb design}
    \label{fig:HoneycombDesign}
\end{figure}

The second pattern we investigated for the design is an honeycomb structure in Figure \ref{fig:HoneycombDesign}. Honeycomb patterns are often used in mirror lightweighting. We choose to implement the pattern with branches of 1.5 mm thickness and 2 mm large and honeycomb nodes located in the centre of the rib structure triangle. The volume added is 443.4 $mm^3$ which is comparable of the one added in the circle design. 

\begin{figure}[htbp]
    \centering
    \includegraphics[width=.8\textwidth]{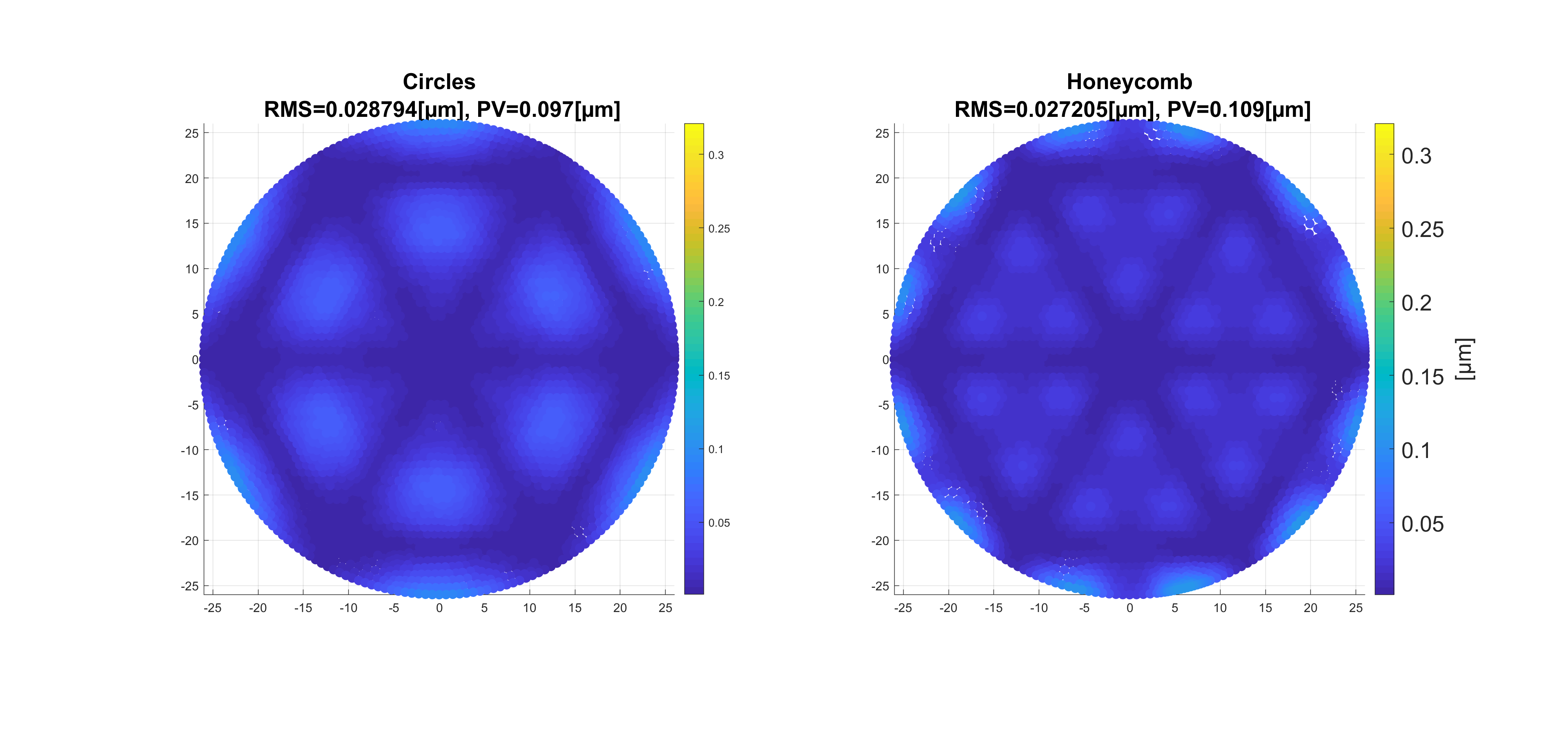}
    \caption{FEA displacement obtained on the circle mirror design and honeycomb mirror design with 0.4 mm surface thickness subject to polishing pressure}
    \label{fig:FEACH}
\end{figure}

Both design show an improvement by a factor three of the optical surface displacements compared to the initial design, in Figure \ref{fig:FEACH}. The imprint pattern is largely reduce on the edges and inside the triangle, the surface form error under polishing is under 30 nm Root Mean Square (RMS).  It is observed that the honeycomb design imprinted higher spatial frequencies upon the surface than the circular design. Circular and honeycomb designs will be printed. In the following section the first set of printing is presented with the circular design. 

\subsection{Prototyping phase}
In this section the prototyping plan is presented. The first set of printing aims to optimise the printing parameters and adjust the design to 3D printing constrains. 

\subsubsection{Material}

We tried two types of material, metal and ceramics, in order to compare the performance of the different printing techniques. Early investigations undertaken by the University of Sheffield identified that they could achieve lower porosity using Ti-6Al-4V (Ti64) than aluminium. Have a low porosity is crucial for polishing steps to avoid pits or cracks. We chose to use Ti64 printed by Electron Beam Melting (EBM) which could provide a fast prototyping. EBM uses electron beam melting techniques to build layer by layer the 3D object. 
To date there have been few investigations using 3D printed ceramics. TNO can provide 3D printed $Al_2O_3$ (alumina), which is stiff and exhibits low porosity. Alumina is printing with stereo-lithography SLA process, which uses photo-polymerisation reaction to build the layer in a liquid resin.  

\begin{table}[htbp]
    \centering
    \begin{tabular}{|l|c|c|}
    \hline
         Properties & Ti64 & Alumina \\
         \hline
         Young Modulus [GPa]& 110 & 360 \\ 
         Poisson Ratio & 0.342 & 0.242 \\
         Density [g/$cm^3$]& 4.43 & 3.9 \\
         \hline
    \end{tabular}
    \caption{Mechanical properties of 3D printing titanium\cite{MURR20101861} \quad  \cite{RAE201710} and alumina\cite{Alumina1} \quad \cite{Alumina2}}
    \label{tab:my_label}
\end{table}{}

\begin{figure}[htbp]
    \centering
    \includegraphics[width=.4\textwidth]{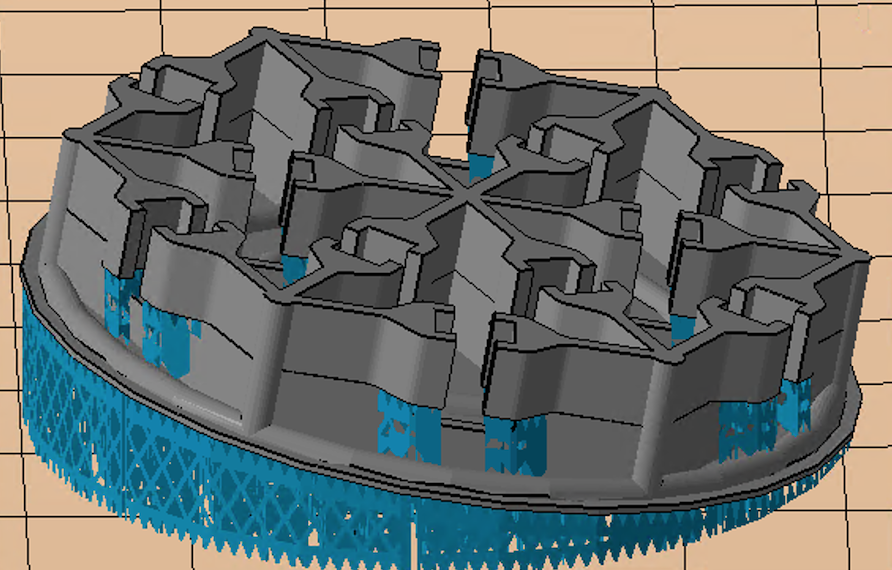}
    \caption{CAD view of the Ti64 model with the additional supports in blue}
    \label{fig:support}
\end{figure}

Before printing, the CAD model is going through the printer software where the design is sliced to defined the laser path on each layer and orientated to provide an optimal quality results at the end. Figure \ref{fig:support} shows the orientation of the Ti64 sample. The optical surface faces down and is inclined thank to supports (in blue) to obtained a better finishing. Additional supports are necessary to build the actuator holders. All the supports are removed after printing. The alumina samples are printing with similar orientation and supports.

\subsubsection{Titanium Printing}
\begin{figure}[htbp]
    \centering
    \includegraphics[width=.4\textwidth]{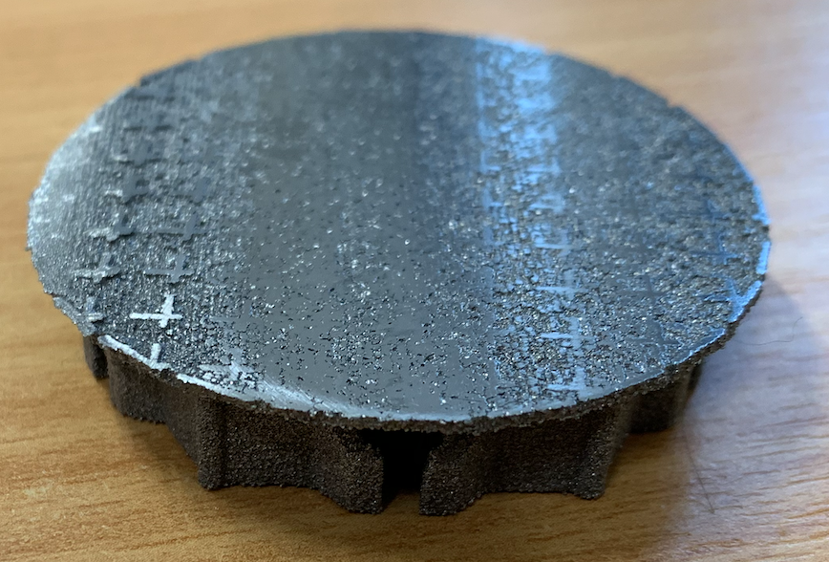}
    \caption{Picture of the Ti64 sample printed by EBM with 0.5 mm surface thickness}
    \label{fig:Tiproto}
\end{figure}

The first set of Ti64 samples shows very rough finishing on the whole surface, with bending on the edges and fused supports on the optical surface. Figure \ref{fig:Tiproto} highlights the bending on the left edge and the support remains corresponding to the cross. The next samples will be built with a higher optical surface thickness to allows more ability to remove the damage caused by the supports. The additional material will be remove during post-processing. From the result of Ti64 printing we decided to do more investigations on the impact of the orientation and the supports during the printing. Porosity location in the sample will be also investigated. 

\subsubsection{Alumina Printing}
\begin{figure}[htbp]
    \centering
    \includegraphics[width = .4\textwidth]{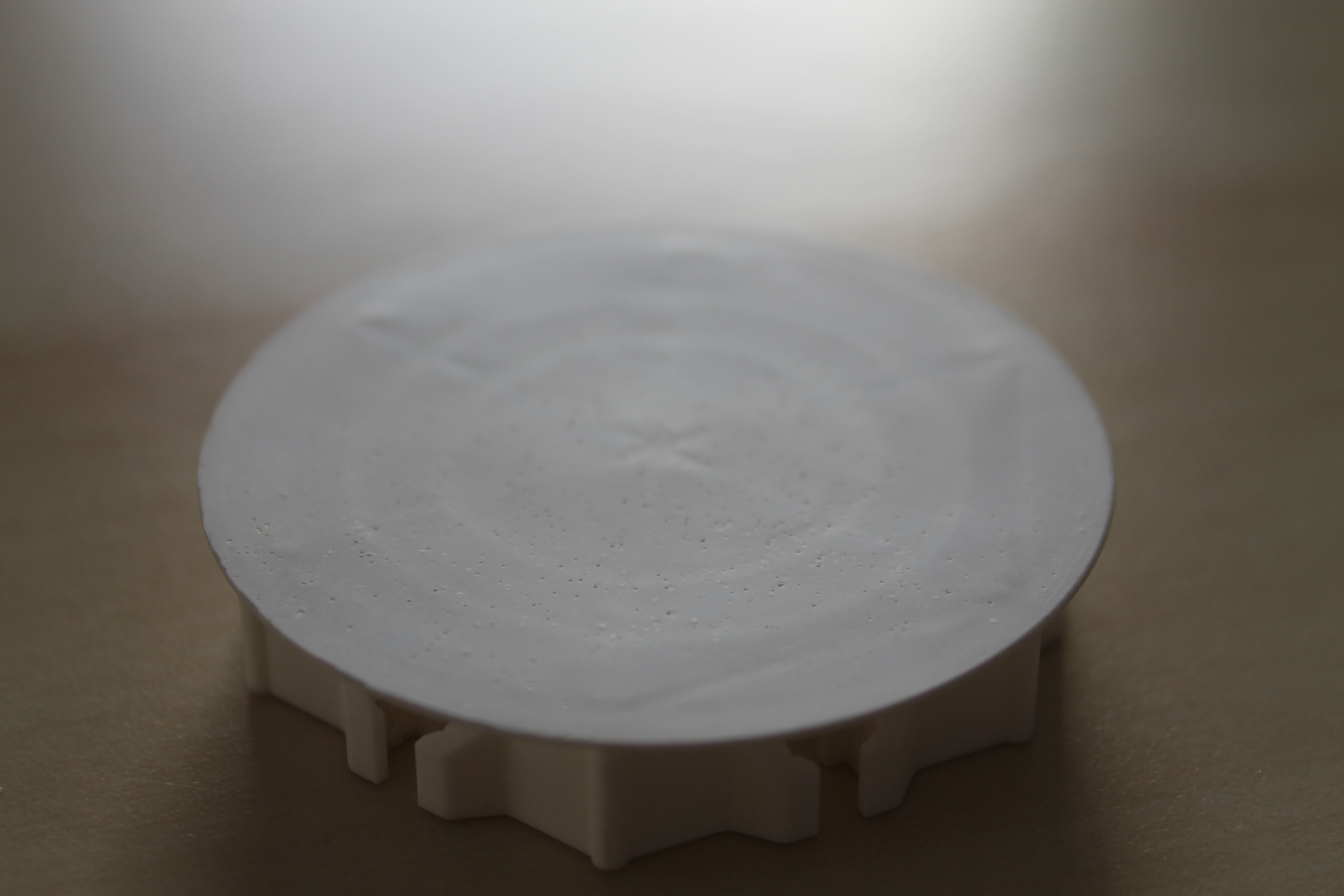}
    \includegraphics[width = .4\textwidth]{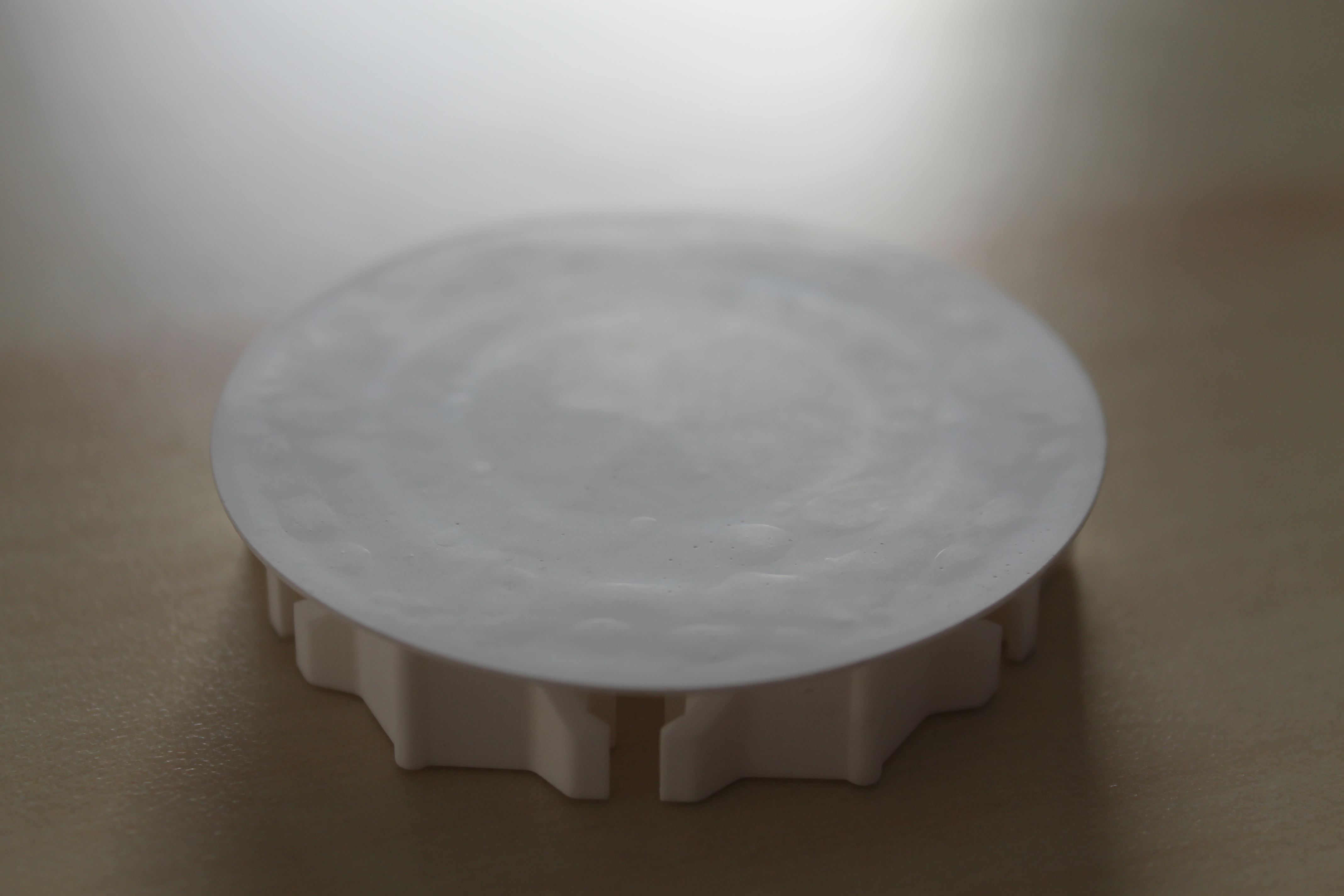}
    \caption{Pictures of two alumina samples printed by SLA with 0.5 mm surface thickness.}
    \label{fig:alumina}
\end{figure}

From Alumina samples the printing shows better finish although there are some pits on the optical surface as well as bubbles, in Figure \ref{fig:alumina}. The optical surface thickness will be increase up to 1 mm to allows more freedom to remove the surface errors in the polishing. 

\subsection{Surface quality measurements}
\begin{figure}[htbp]
    \centering
    \includegraphics[width=.55\textwidth]{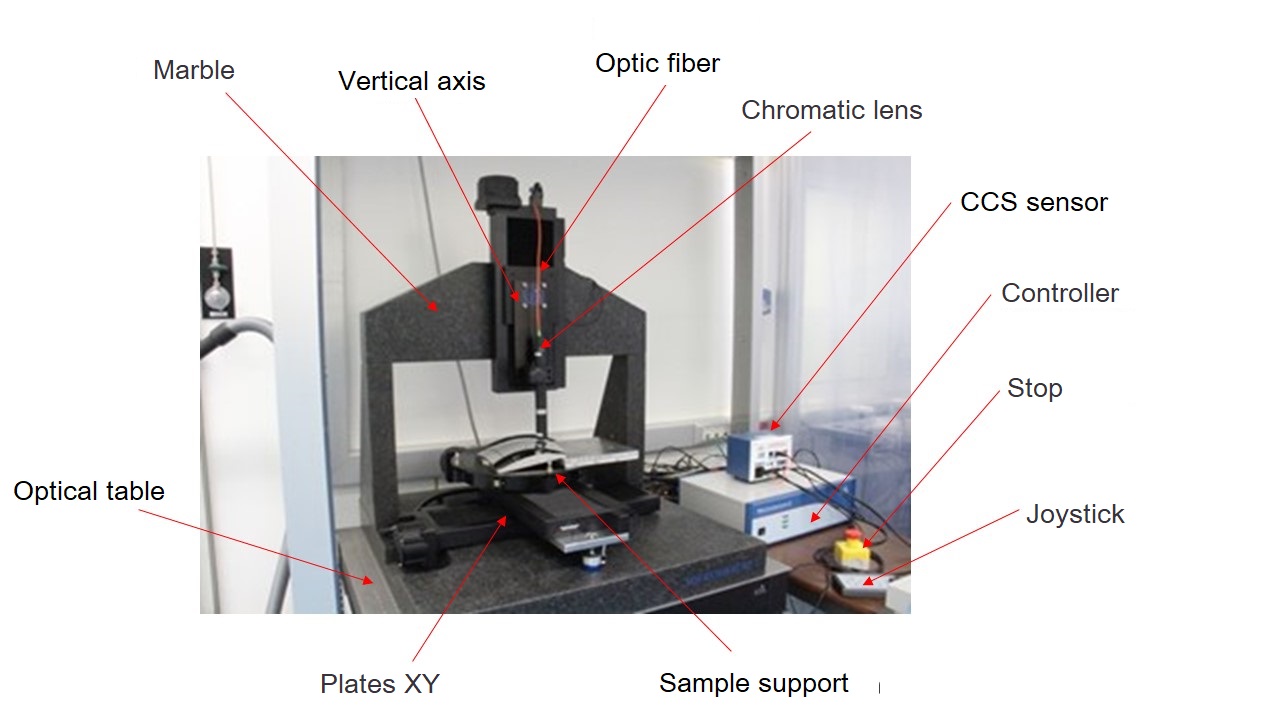}
    \includegraphics[width=.4\textwidth]{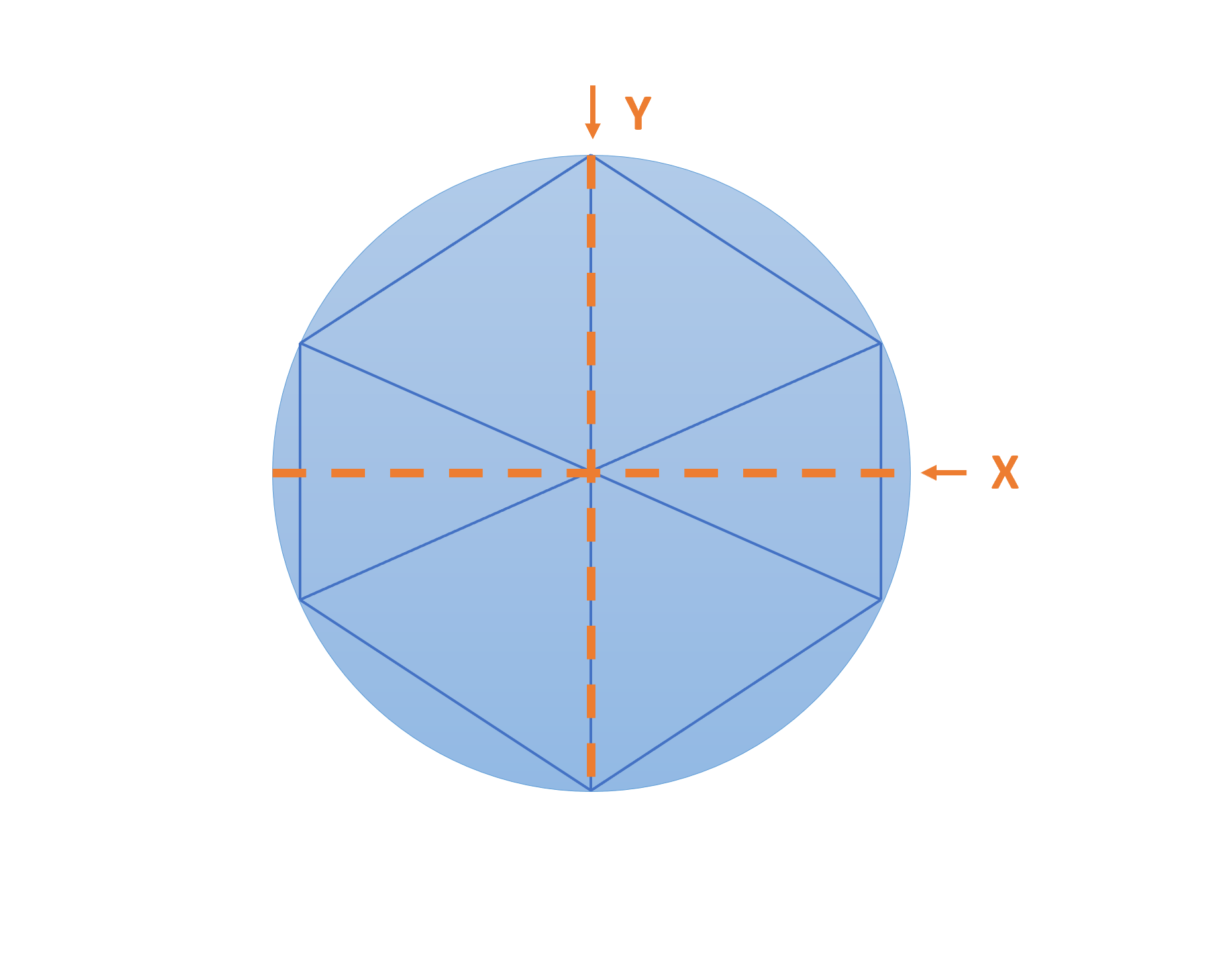}
    \caption{Left: Picture of the STIL profilometry instrument. Right: Sketch of the sample with the measured profiles X and Y}
    \label{fig:Profile}
\end{figure}

\begin{figure}[htbp]
    \centering
    \includegraphics[width=1\textwidth]{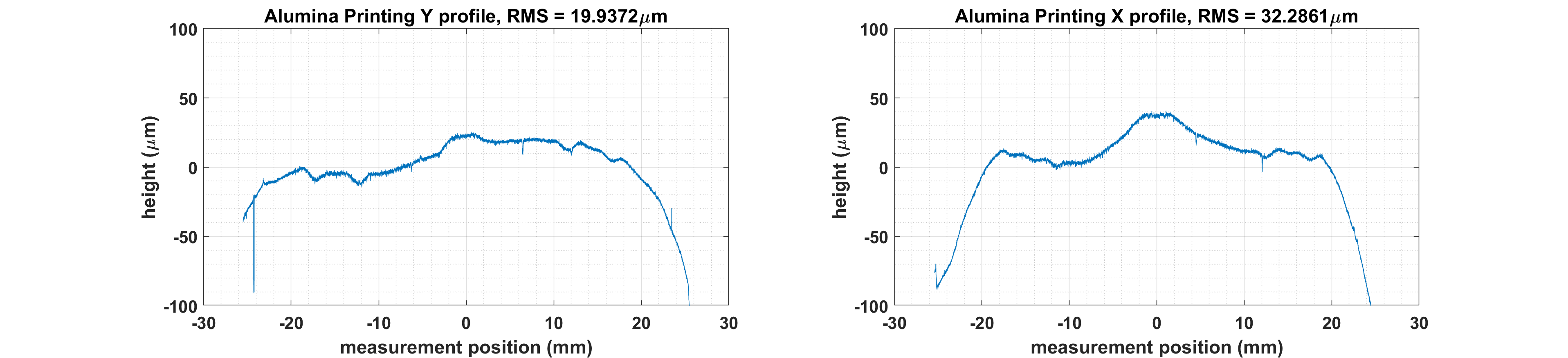}
    \caption{Profilometry measurement on alumina 1 mm surface thickness after printing on the Y axis (left) and X axis (right)}
    \label{fig:ProfiloP}
\end{figure}

\begin{figure}[htbp]
    \centering
    \includegraphics[width=1\textwidth]{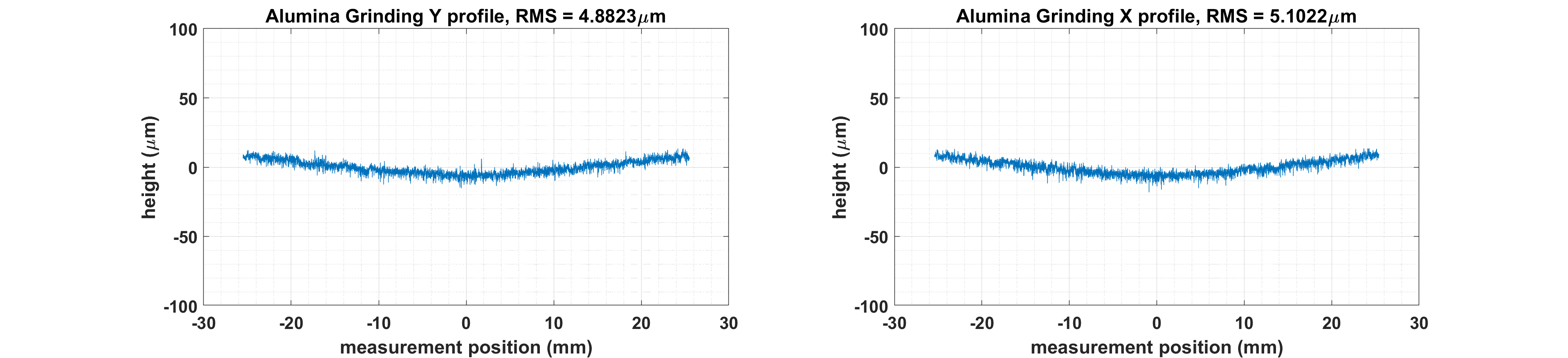}
    \caption{Profilometry measurement on alumina 1 mm surface thickness after flat rough grinding}
    \label{fig:ProfiloG}
\end{figure}

Profilometry measurements were performed on the alumina mirror with 1 mm optical surface thickness in a ISO 5 cleanroom with a STIL profilometer as shown in Figure \ref{fig:Profile} \textit{left}. The measures of the optical surface are taken on the Y axis, along a rib structure and on the X axis going through two triangle patterns, as shown in Figure \ref{fig:Profile} \textit{right}. Two measurements are performed, one immediately after printing (Figure \ref{fig:ProfiloP}) and the second after flat grinding post-processing (Figure \ref{fig:ProfiloG}).

After printing in Figure \ref{fig:ProfiloP}, both profiles show bending on the edges. In the X profile two additional bending zones are observed located in the middle of the triangle pattern resulting of the the lack of supports. The same magnitude of distortion is not observed in the Y profile. The printing affects the flatness of the optical surface by introduce a sag between the rib structure which is mainly caused by the sample orientation during printing. This effect also explains the the bump in the centre of the sample. 

\begin{figure}[htbp]
    \centering
    \includegraphics[width=.4\textwidth]{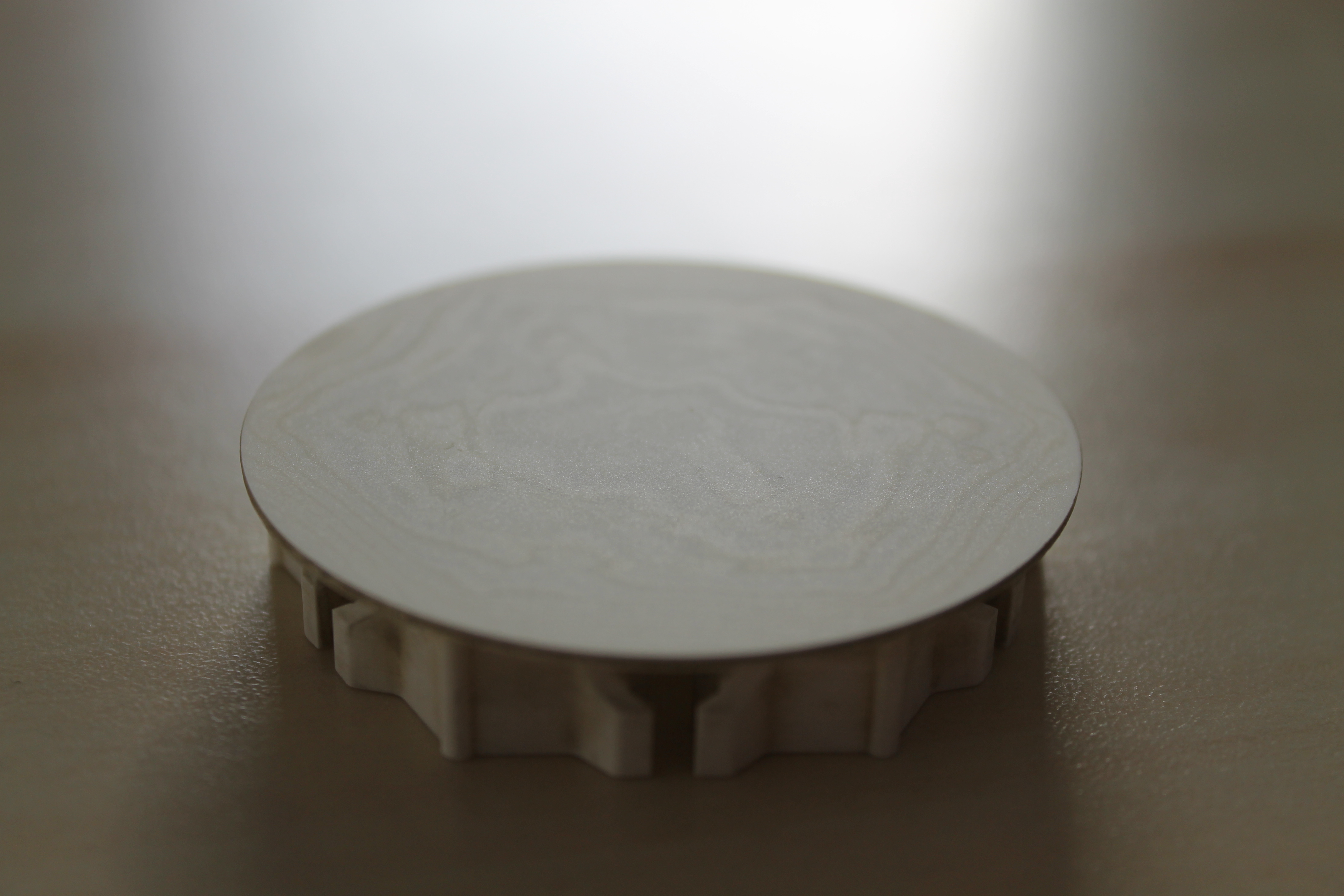}
    \caption{Photo of a 1 mm thick alumina sample after flat grinding}
    \label{fig:AlPoli}
\end{figure}

A first try of rough flat grinding is performed on the Alumina sample with corundom polishing compound. Profilometry measurements are then performed along the same profiles and the results are shown in Figure \ref{fig:ProfiloG}. The surface errors from the printing can be easily removed with grinding. However the high grain size increase the roughness. Although the profilometry measurement are quite promising, the visual aspect of the mirror is more concerning. During polishing pits appeared on the surface and we can clearly see the layer of printing with a hexagonal/star shape, in Figure \ref{fig:AlPoli}. In the profilometry measurement there are too much roughness to highlight this visual effect. 

\subsection{Future printing}

A future set of printing is on going to print both circle design, in Figure \ref{fig:CircleDesign} and honeycomb design in Figure \ref{fig:HoneycombDesign}. For each design, four samples will be printed in alumina with a surface thickness of 1 mm. Eights models are printed in Ti64 with a 2 mm surface thickness. Four of the samples are subjected to Hot Isostatic Pressing (HiP) at Sheffield University. HiP is a post-processing where the samples are elevated to high temperature in an isostatic high gas pressure environment. This process aims to reduce porosity and increase the density of the material. We are expecting a better polishing response with this additional post-processing.

\begin{table}[htbp]
    \centering
    \begin{tabular}{|c|c|c|c|c|}
    \hline
       Designs & Alumina printing & Titanium printing & Titanium printing + HiP & Total\\
       \hline
        Circle & 4 & 4 & 4 & 12 \\
        Honeycomb & 4 & 4 & 4 & 12\\
        \hline
    \end{tabular}
    \caption{Resume of the future printing samples}
    \label{tab:future}
\end{table}{}

In order to characterise the quality of the different printing methods, all 24 samples (Ti64 and alumina) will be measured. First to validate the dimensional properties (diameter, surface thickness and height). Then to characterise the surface quality with profilometry measurements and compare the different 3D printing techniques and materials. 
All the mirrors will be subject to the same polishing method. They will be flat polishing up to 0.4mm surface thickness. 
A final profilometry measurement will compared the results on the surface quality in terms of surface errors and roughness for all the samples. Thus we could validate the simulations and compare the behaviour under polishing of the different materials. 

\section{Stress Mirror Polishing}

\begin{figure}[htbp]
    \centering
    \includegraphics[width=.5\textwidth]{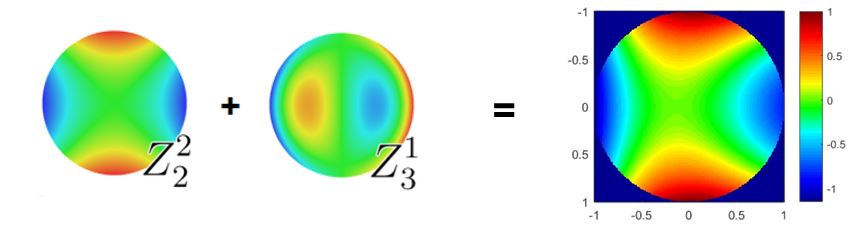}
    \caption{Zernike decomposition of an off-axis parabola}
    \label{fig:ZOAP}
\end{figure}

3D printing techniques removes the traditional manufacturing constraints and therefore opens a new range of design possibilities. In this section we are exploring a new mirror thickness distribution capable of producing off-axis parabolas (OAPs) using Stress Mirror Polishing technique (SMP). The design investigation is performed in parallel to the WFIRST project where LAM is involve in the fabrication of the coronagraph OAPs \cite{10.1117/12.2312633}. OAPs are used as a relay, they are located between the other optical components to relay the beam between the focal plane to the pupil plane. The surface of each OAP is required superpolishing to minimise the introduction of wave-front errors in the beam transmission. An OAP can be defined with Zernike polynomials \cite{Lubliner:80} by combination of astigmatism 3x and coma 3x, as shown in Figure \ref{fig:ZOAP}. 

\subsection{Stress Polishing of Astigmatism mirror}

Stress polishing is a technique developed by the German astronomer Bernhard Schmidt in the 1930s \cite{Lemaitre:72}. The process consists in applying forces through a warping harness on the substrate while polishing the part as a spherical surface. During this step the deformation created by the warping harness is imprinted on the substrate surface. After the polishing phase, the warping harness is removed and the substrate comes back to its initial position. To use this process the warping harness must generate a warping function equal to the inverse of the required final shape and the forces applied must be under yield strength of the material. This technique provides a very high surface quality with an easy manufacturing and is perfectly suited for high contrast imaging. Toric mirrors has been manufactured at Laboratoire d'Astrophysique de Marseille LAM for the Spectropolarimeter SPHERE on the VLT \cite{refId0} and for High Contrast bench HiCat \cite{10.1117/12.2056694}. 

\begin{figure}[htbp]
    \centering
    \includegraphics[width = .3\textwidth]{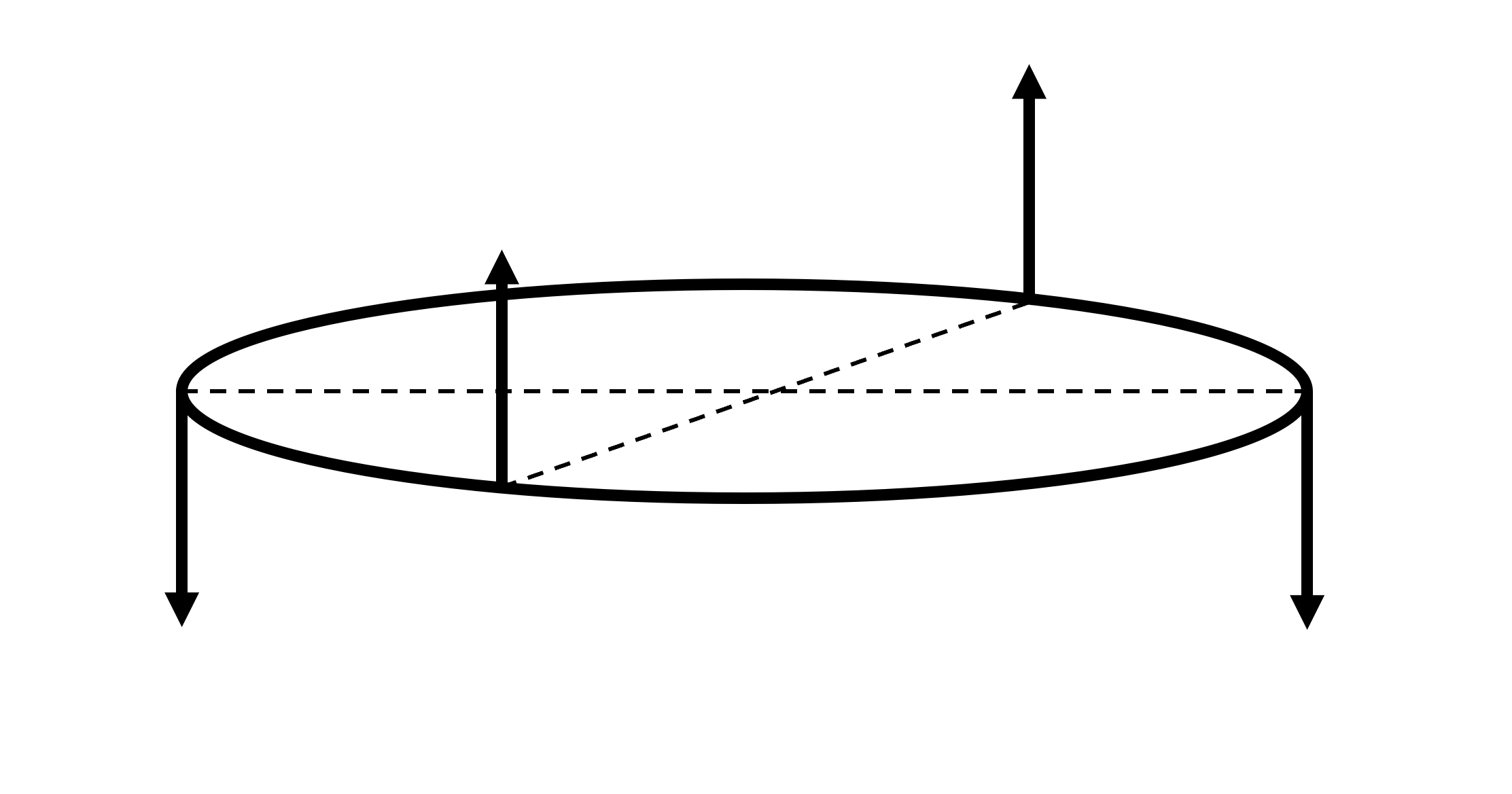}
    \includegraphics[width = .3\textwidth]{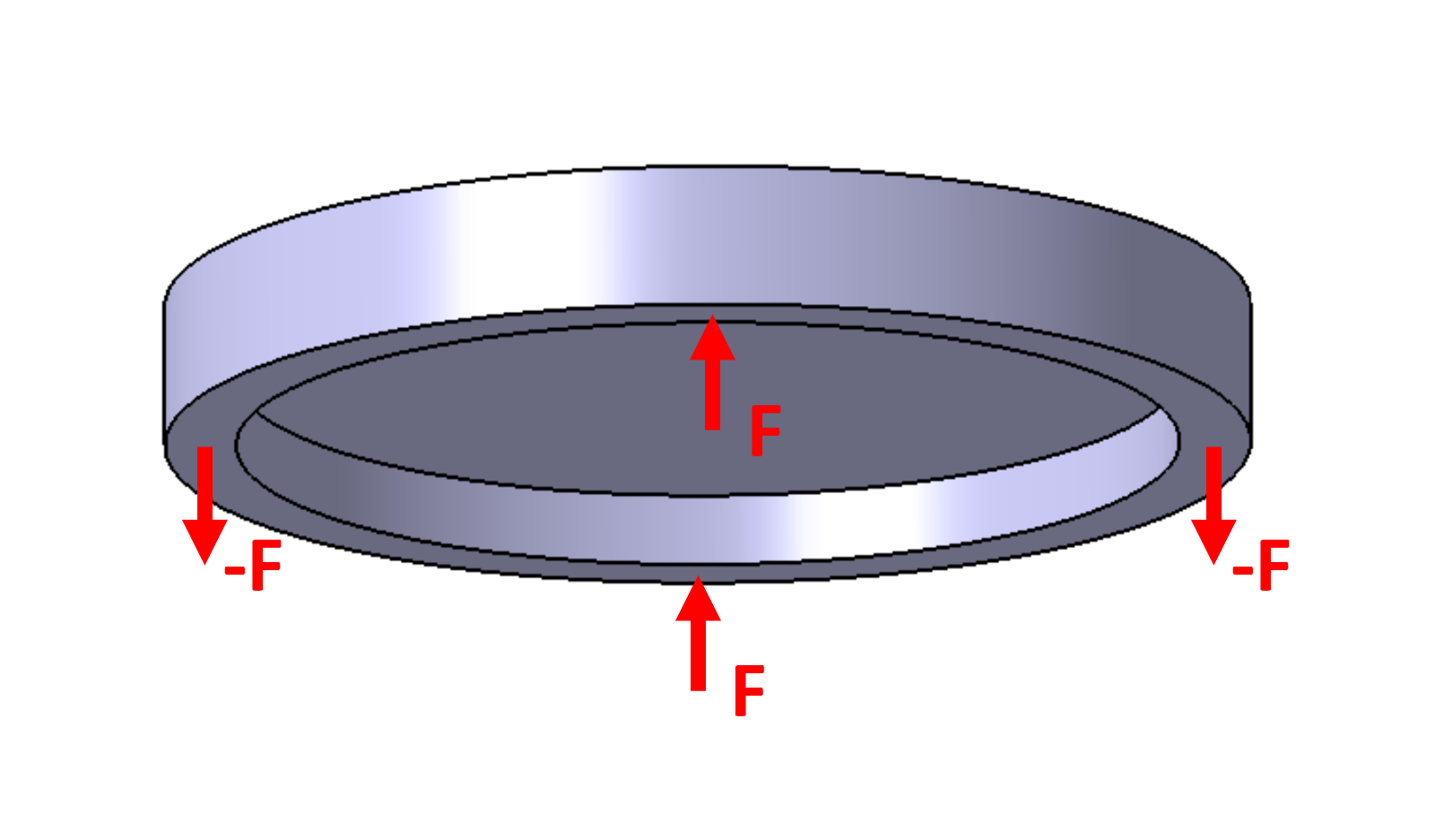}
    \includegraphics[width = .3\textwidth]{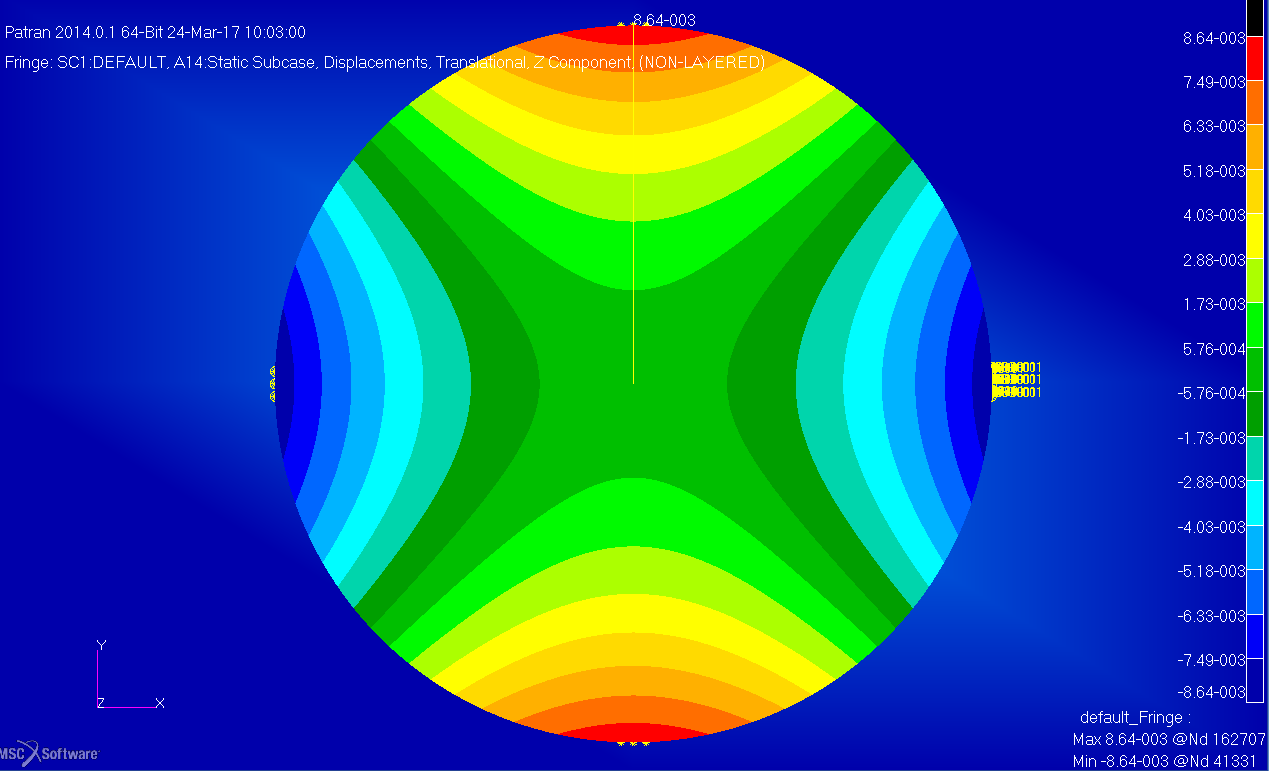}
    \caption{Left: Sketch of the combination of forces needed to generate astigmatism. Middle: CAD view of the mirror design with the ring and the force application in red. Right: Displacement fringes on the surface after FEA corresponding to astigmatism 3x. }
    \label{fig:Astig}
\end{figure}

In the case of toric mirror, two pairs of opposite forces are applied on a ring at the back side of the mirror, in Figure \ref{fig:Astig}. This combination of force is simple to achieve with one actuator device or two screws system. The deformation is simulated with FEA. Figure \ref{fig:Astig} right shows the displacement fringes of the deformation, we can clearly identify the astigmatism pattern. Then Zernike decomposition is used to characterise the surface shape. 

\subsection{OAP design and prototyping}
Starting from the design of toric mirror create by E. Hugot et al. (2012) \cite{refId0} we performed a parametric study on the thickness distribution to generate the OAP profile via astigmatism 3x and coma 3x with the same combination of forces (Figure \ref{fig:Astig}) - this study will be presented in M. Roulet et al. (2020).

\begin{figure}[htbp]
    \centering
    \includegraphics[width = .4\textwidth]{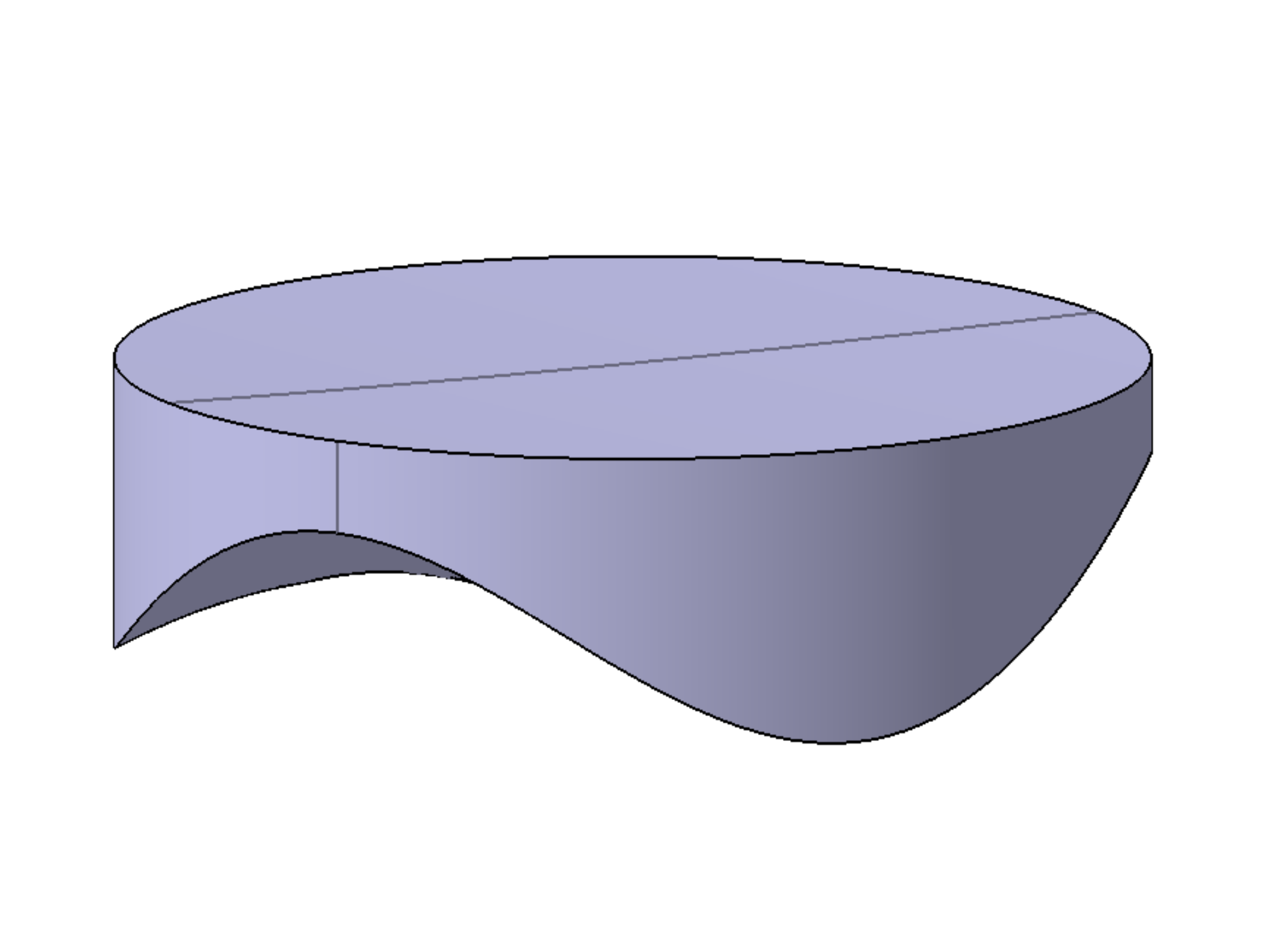}\qquad\qquad\quad
    \includegraphics[width = .3\textwidth]{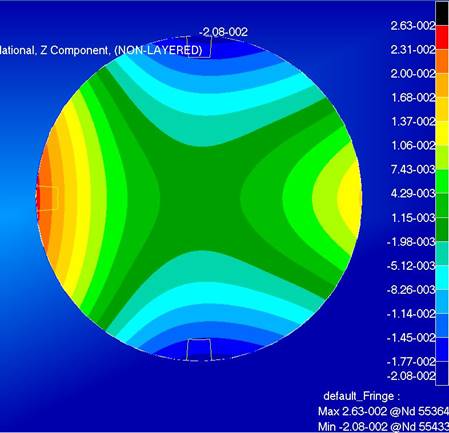}
    \caption{Left: CAD view of the mirror design with the thickness distribution generation OAP. Right: Displacement fringes on the surface after FEA corresponding to astigmatism 3x and coma 3x.}
    \label{fig:OAPsimu}
\end{figure}

The final design is shown in Figure \ref{fig:OAPsimu} left. The thickness distribution is composed by three wave-forms. On the right side of Figure \ref{fig:OAPsimu} the displacements on the optical surface show the OAP shape. The Zernike decomposition confirmed that the new thickness distribution generate Astigmatism 3x and Coma 3x with very low residuals. 

\begin{figure}[htbp]
    \centering
    \includegraphics[width = .3\textwidth]{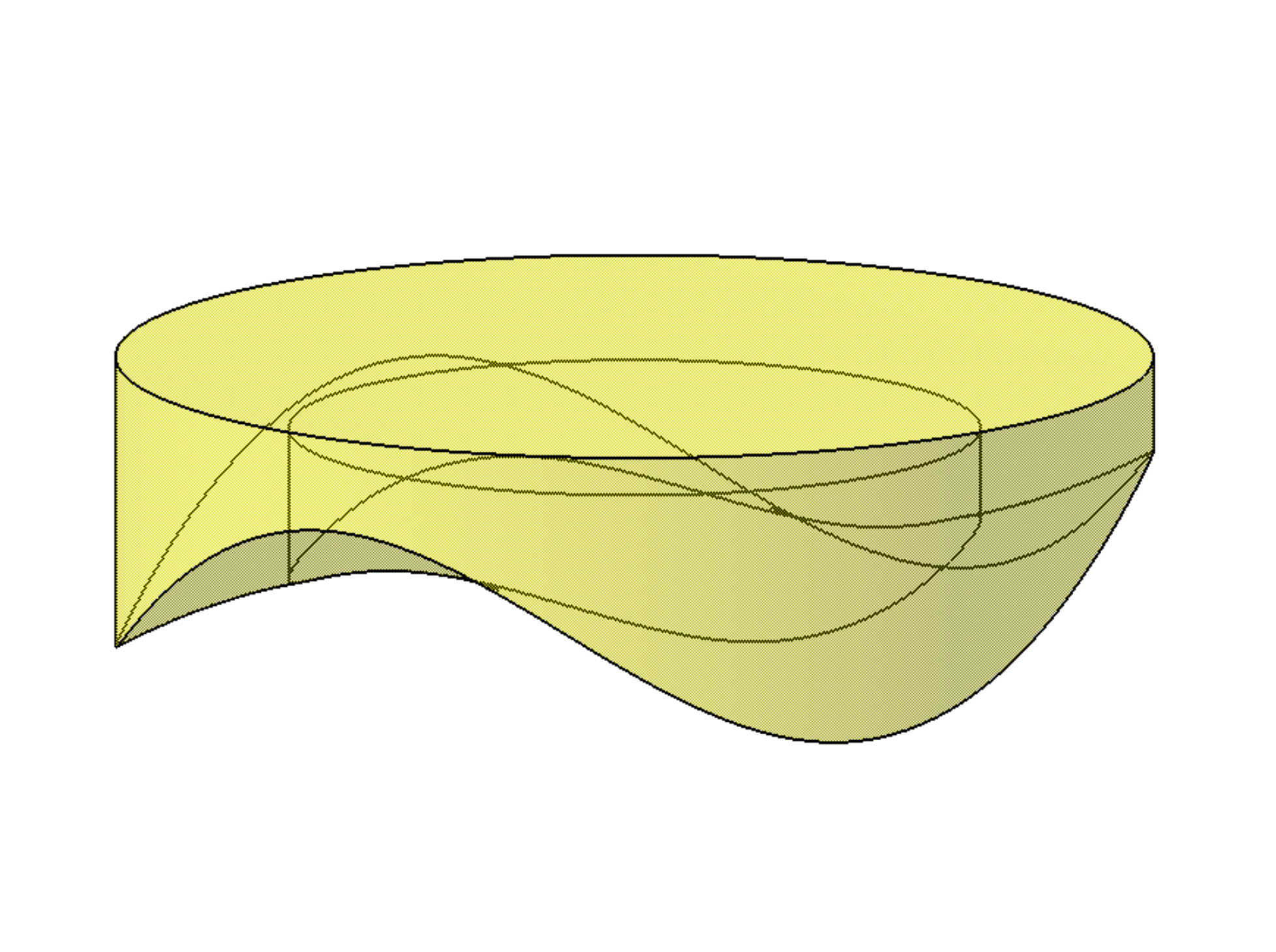}
    \includegraphics[width = .3\textwidth]{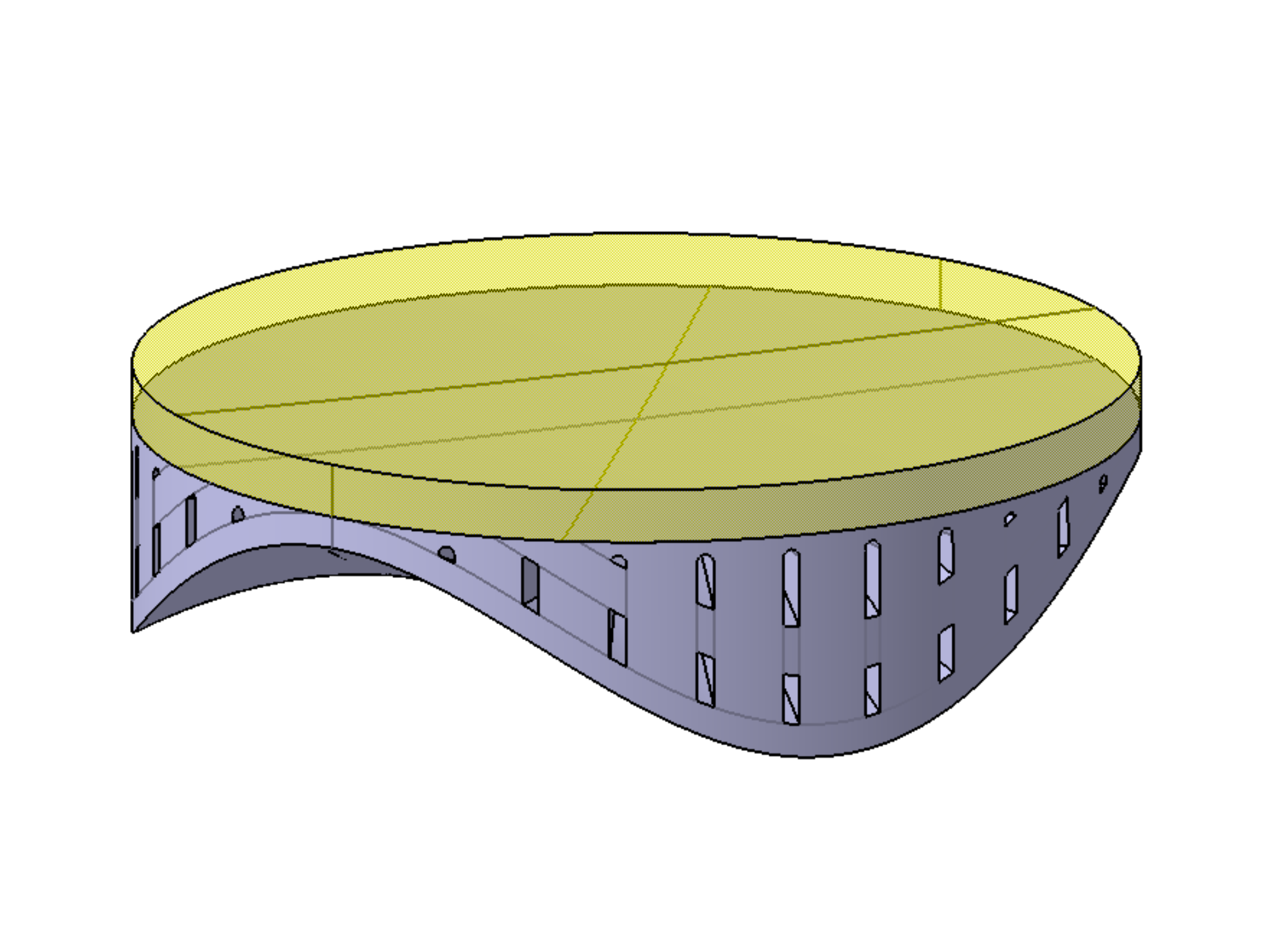}
    \includegraphics[width = .3\textwidth]{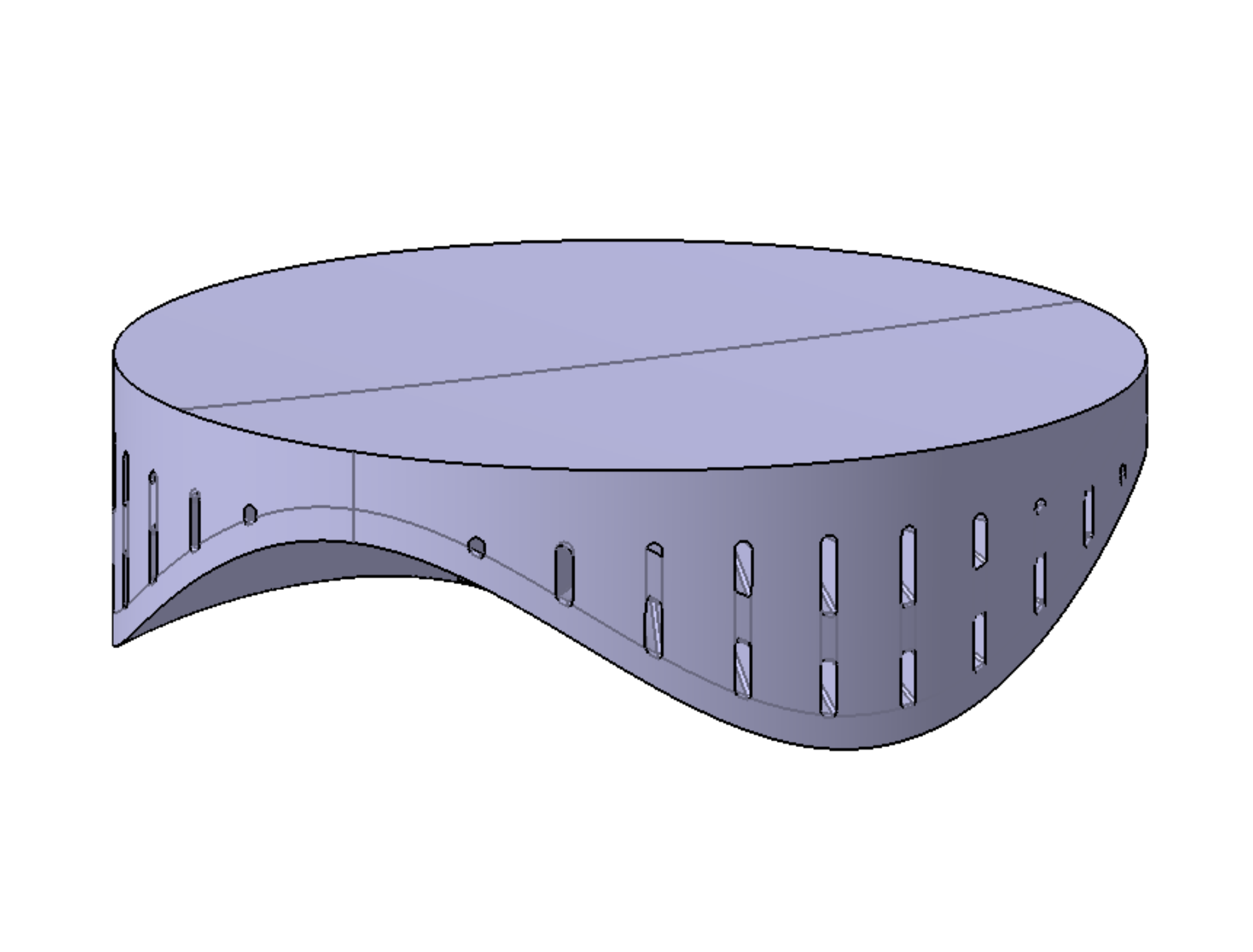}
    \caption{CAD views of the three prototypes. Left: the full Zerodur Prototype. Middle: the two-part assembly prototype. Right: the full 3D printed prototype.}
    \label{fig:OAPproto}
\end{figure}

From the design in Figure \ref{fig:OAPsimu} left, three prototypes are manufactured, they are presented in Figure \ref{fig:OAPproto}. The objective is to validate experimentally the new thickness distribution and how 3D printing can be applied within stress polishing. 

The first prototype in Figure \ref{fig:OAPproto} \textit{left} is traditional manufacturing from a Zerodur milled blank. This prototype has been polished and deformed, the results are presented in M. Roulet et al (2020). 

Two further prototypes are manufactured by 3D printing: a Two-Part Assembly (TPA) prototype, in Figure \ref{fig:OAPproto} \textit{middle} and a full 3D printed prototype, in Figure \ref{fig:OAPproto} \textit{right}. A ceramic, cordierite, is used for its mechanical properties close to Zerodur. The Young's modulus of cordierite is 140 GPa and its coefficient of thermal expansion (CTE) is very low, between 0.2 and 0.9 $10^6 K^-1$\cite{Cordierite}. This properties made a good candidate to replace Zerodur in future mirror fabrication. 
Lightweighting the thickness distribution is mandatory to ensure the correct material properties after printing; a maximum thickness of 5 mm is required to allow the material to cure appropriately. The TPA prototype in Figure \ref{fig:OAPproto} \textit{middle} is a combination of the two firsts. The thickness distribution is printed in cordierite and a milled blank of Zerodur is bonded on the top. 

\subsection{Deformation and measurement}
\begin{figure}[htbp]
    \centering
    \includegraphics[width = .41\textwidth]{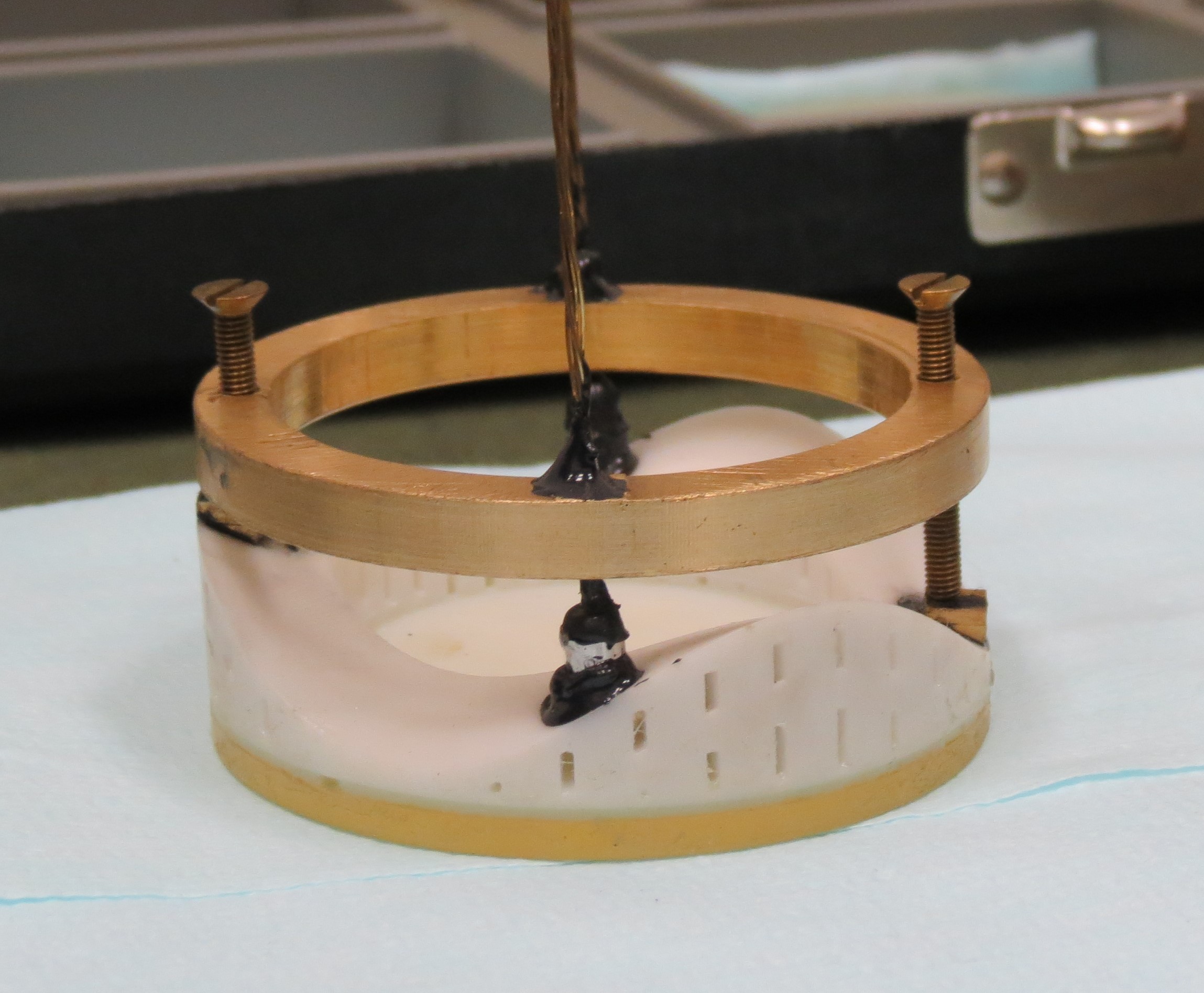}
    \includegraphics[width = .45\textwidth]{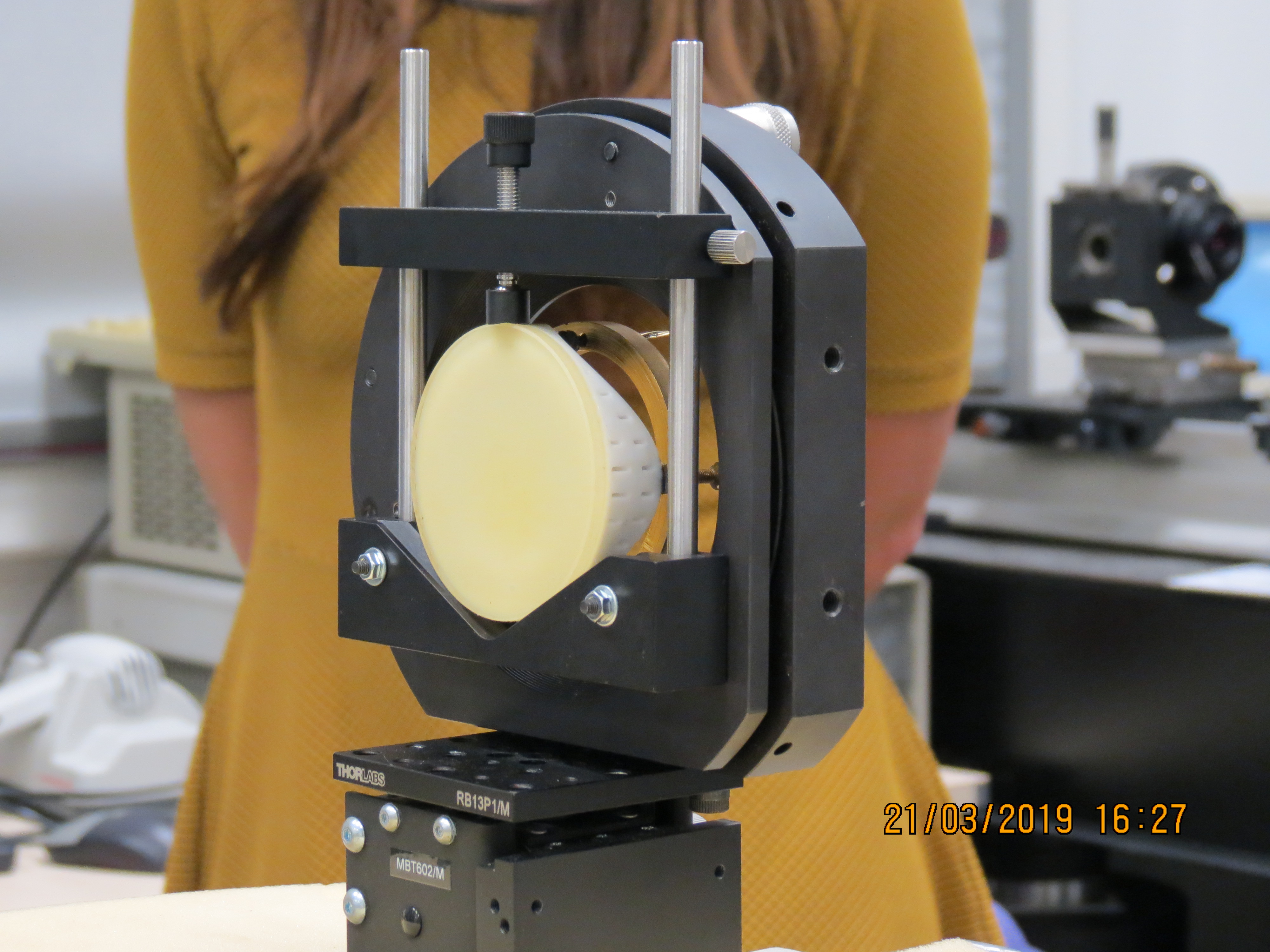}
    \caption{Pictures of the TPA prototype. Left: with its warping harness. Right: The prototype is mounting on the interferometry bench.}
    \label{fig:OAP7 assembly}
\end{figure}

The TPA prototype has been polished spherically up to obtained a reflective surface and then assembled to its warping harness in Figure\ref{fig:OAP7 assembly} \textit{left}. The warping harness is composed of a ring attached to two screws for the pushing forces and two wires for the pulling forces. In Figure \ref{fig:OAP7 assembly} \textit{right}, the TPA prototype is mounted on the interferometry bench to measure the deformation. 

Comparison between the simulated deformation and the interferometric measurement is shown in Table \ref{tab:Result}.
During the measurement, the deformation is adjust to the simulated astigmatism 3x magnitude. To evaluate the suitability of the deformation we introduce the ratio $R_{Astig/Coma}$ which represents the magnitude of coma 3x normalised by astigmatism 3x.
The simulated and measured $R_{Astig/Coma}$ are closed, with 1\% difference thus from this experimentation we validated the simulation technique. The residuals are computed up to the 21st Zernike coefficient and removing tip, tilt, piston, astigmatism 3x and coma 3x. The experimental measurement shows higher residuals than the ones simulated. This difference is mainly due to the interfaces between the Zerodur blank and the printed thickness distribution, but also from the assembly of the warping harness and the bonding pads force application. Indeed in the simulation the forces are perfectly located and normal to the optical surface while in the experiment pads and glue are added to create the force application surfaces.

\begin{table}[htbp]
    \centering
    \begin{tabular}{|c|c|c|}
    \hline
       Zernike polynomials RMS [nm] & Simulated deformation & Measured deformation  \\
       \hline
        Astig 3x & 1316 & 1316.4\\
        Coma 3x & 153 & 129.1\\
        Residuals & 71 & 84.0\\
        \hline
        $R_{Astig/Coma}$ & 11.34\% & 10.2\%\\
        \hline
    \end{tabular}
    \caption{Comparison between simulated deformation and the measured deformation}
    \label{tab:Result}
\end{table}{}

\section{Conclusion}

This paper presented two prototypes using 3D printing for astronomy mirror fabrication. We highlight the advantages of 3D printing to solve the issue of interface error and design constraints from traditional manufacturing. 
In the case of thin deformable mirror a new set of 24 samples will be printed and will be polished and measured to characterise their behaviour under polishing. Regarding the stress polishing technique, we prove that is it possible to generate off-axis parabola by by the introduction of a new thickness distribution on the back of the mirror via 3D printing, while maintaining ease of assembly and polishing. From these first two promising proof-of-concepts, further investigation both on designs and prototyping are needed to develop the use of 3D printing astronomical mirrors. 

\acknowledgments 
 
This project has received funding from the European Union's Horizon 2020 research and innovation programme under grant agreement No 730890. This material reflects only the authors views and the Commission is not liable for any use that may be made of the information contained therein. 
The authors also would like to acknowledge the European commission for funding this work through the Program H2020-ERC-STG-2015 – 678777 of the European Research Council.

\bibliography{report} 
\bibliographystyle{spiebib} 

\end{document}